\documentstyle[aas2pp4]{article}

\begin{document}
\title{HEAVY ELEMENT ABUNDANCES IN BLUE COMPACT GALAXIES}
\author{Yuri I. Izotov}
\affil{Main Astronomical Observatory, Ukrainian National Academy of Sciences,
Goloseevo, Kiev 252650, Ukraine \\ Electronic mail: izotov@mao.kiev.ua}
\and
\author{Trinh X. Thuan}
\affil{Astronomy Department, University of Virginia, Charlottesville
VA 22903 \\ Electronic mail: txt@virginia.edu}

\begin{abstract}

We present high-quality ground-based spectroscopic 
observations of 54 supergiant H II regions in 50 low-metallicity blue
compact galaxies with oxygen abundances 12 + log O/H between 7.1 and  8.3. We 
use the data to determine abundances for the
elements N, O, Ne, S, Ar and Fe. We also analyze {\sl Hubble Space 
Telescope} ({\sl HST}) Faint Object Spectrograph archival spectra of 
10 supergiant H II regions to derive C and Si abundances in a subsample of 7  
BCGs. The main result of the present study
is that {\it none} of the heavy element-to-oxygen abundance ratios
studied here (C/O, N/O, Ne/O, Si/O, S/O, Ar/O, Fe/O) depend on oxygen
abundance for BCGs with 12 + log O/H $\leq$ 7.6 ($Z$ $\leq$ $Z_\odot$/20). This 
constancy implies that all these heavy elements have a primary origin and are 
produced by the same massive ($M$ $\geq$ 10 $M_\odot$) stars responsible for O 
production. The dispersion of the C/O and N/O ratios in these galaxies is found 
to be remarkably small, being only $\pm$ 0.03 dex and $\pm$ 0.02 dex respectively. 
This very small dispersion is strong evidence against any time-delayed 
production of C and primary N 
in the lowest-metallicity BCGs (secondary N production is negligible at these low 
metallicities). The absence of a time-delayed production of C and N 
is consistent with the scenario that 
galaxies
with 12 + log O/H $\leq$ 7.6 are undergoing now their {\it first} burst
of star formation, and that they are therefore {\it young}, with ages not
exceeding 40 Myr. If very low metallicities BCGs are indeed young, this 
 would argue against the commonly held belief 
that C and N are produced  
by intermediate-mass (3 $M_\odot$ $\leq$ $M$ $\leq$ 
9 $M_\odot$) stars at very low 
metallicities, as these stars would not have 
 yet completed their evolution in these lowest metallicity galaxies.
In higher metallicity BCGs (7.6  $<$ 12 + log O/H $<$ 8.2), the 
Ne/O, Si/O, S/O, Ar/O and Fe/O abundance ratios retain the same constant value 
they had at lower metallicities. By contrast, there is an increase of the C/O 
and N/O ratios along with their dispersions at a given O. We interpret this 
increase as due to the additional contribution
of C and primary N production in intermediate-mass stars, 
on top of that by high-mass stars. The above results lead to the following 
timeline for galaxy evolution:  a) all objects with 12 + log O/H $\leq$ 7.6 began 
to form stars less than 40 Myr ago; b) after 40 Myr, all galaxies have evolved 
so that 12 + log O/H $>$ 7.6; c) by the time intermediate-mass stars have
evolved and released their nucleosynthetic products (100--500 Myr), 
all galaxies have become enriched to 7.6 $<$ 12 + log O/H $<$ 8.2.  
The delayed release of primary N at these metallicities greatly increase
the scatter in the N/O abundance ratio; d) later, when galaxies get enriched to
12 + log O/H $>$ 8.2, secondary N production becomes important.

BCGs show the same O/Fe 
overabundance with respect to the Sun ($\sim$ 0.4 dex) as galactic halo stars, 
suggesting the same chemical enrichment history. 
We compare heavy elements yields derived from the
observed abundance ratios with theoretical yields for massive stars
and find general good agreement. However, the theoretical models are unable
to reproduce the observed N/O and Fe/O abundance ratios. Further theoretical 
developments are necessary, in particular to solve the problem of primary 
nitrogen production in low-metallicity massive stars.

 We discuss the apparent discrepancy between the N/O abundance
ratios measured in BCGs and those in high-redshift damped Ly$\alpha$ galaxies, 
which are up to one order of magnitude smaller. 
We argue that this large discrepancy may arise from the unknown  
physical conditions of the gas responsible for the metallic absorption lines in 
high-redshift damped Ly$\alpha$ systems. While it is widely assumed
that the absorbing gas is neutral, we propose that it could be ionized.
In this case, ionization correction factors can boost the N/O ratios in 
damped Ly$\alpha$ galaxies into the range of those measured in BCGs.

\end{abstract}


\section {INTRODUCTION}

   The study of the variations of one chemical element relative 
to another is crucial for our understanding of the chemical evolution
of galaxies and for constraining models of stellar nucleosynthesis and
the shape of the initial mass function. Blue compact galaxies (BCGs)
are ideal objects in which to carry out such a study. They are
low-luminosity ($M_B$ $>$ --18 mag) dwarf systems 
undergoing intense bursts of star formation which give birth to a large
number (10$^3$ -- 10$^4$) of massive stars in a compact region, which
ionize the interstellar medium, producing high-excitation supergiant
H II regions, and enrich it with heavy elements (see Thuan 1991 for a
review). The optical spectra of these H II regions show strong narrow
emission lines superposed on a stellar continuum which is rising toward
the blue, allowing abundance determination of such heavy elements as 
nitrogen, oxygen, neon, sulfur, argon and iron. Recently, the 
{\sl Hubble Space Telescope} ({\sl HST}) has permitted
the determination of carbon and silicon abundances from emission lines
in the UV range (Garnett et al. 1995ab; Thuan, Izotov \& Lipovetsky 1996;
Garnett et al. 1997; Kobulnicky et al. 1997; Kobulnicky \& Skillman
1998; Thuan, Izotov \& Foltz 1998).

   These studies are particularly important for understanding the 
evolutionary status of extremely low-metallicity BCGs. The debate has been 
raging for decades, ever since the discovery paper by Sargent \& Searle (1970), 
on whether BCGs are truly nearby young dwarf galaxies undergoing
one of their first bursts of star formation, or the present starburst is 
occurring within an older galaxy. The abundances of heavy elements
in these galaxies ranging between $Z_\odot$/50 and $Z_\odot$/3 make them 
among the least chemically evolved galaxies in the universe. However, 
subsequent photometric studies have shown that the majority of blue compact 
galaxies possess an old stellar population. Hence they have experienced star 
formation episodes in the past and are not young. 
Loose \& Thuan
(1985) found that $\sim$ 95\% of the BCGs in their sample exhibit an underlying 
extended red low surface brightness component, on which are superposed the high
surface brightness blue star-forming regions. Later CCD surveys of BCGs have
confirmed and strengthened that initial result (Kunth, Maurogordato \& Vigroux
1988; Papaderos et al. 1996; Telles \& Terlevich 1997). However, there are
at least two known objects with extremely young stellar populations. They are
the two most-metal deficient galaxies known, I Zw 18 ($Z_\odot$/50, Searle \& 
Sargent 1972) and SBS 0335--052.
{\sl Hubble Space Telescope} ({\sl HST}) imaging of I Zw 18 to $V$ $\sim$ 26 mag 
by Hunter \&
Thronson (1995) suggests that the stellar population is dominated by young
stars and that the colors of the underlying diffuse component are consistent 
with those from a sea of unresolved B or early A stars, with no evidence
for stars older than $\sim$ 10$^7$ yr. The BCG SBS 0335--052 was first shown by
Izotov et al. (1990) to possess an extraordinarily low metallicity, equal to 
1/41 of the Sun's metallicity (Melnick, Heydari-Malayeri \& Leisy 1992). 
{\sl HST}
$V$ and $I$ imaging of this galaxy (Thuan, Izotov \& Lipovetsky 1997) reveals 
extraordinarily blue $(V-I)$ colors (between 0.0 and 0.2 mag), not only in the 
region of current star formation, but also in the extended ($\sim$ 4 kpc in 
size) low surface brightness underlying component. The blue low intensity 
emission has been shown by Izotov et al. (1997a) and Papaderos et al. (1998) to 
be the 
combined effect of emission from ionized gas (contributing $\sim$ 1/3 of the 
total flux) and from a very young stellar population (contributing the remaining 
$\sim$ 2/3 of the emission), not older than $\sim$ 10$^8$ yr. There is further 
observational evidence for the extremely young age of SBS 0335--052. Thuan \& 
Izotov (1997) have argued from {\sl HST} UV spectrophotometry that the large 
(some 64 kpc in size) and massive ($\sim$ 10$^9$ $M_\odot$) H I cloud associated 
with the BCG (Pustilnik et al. 1998) is made of pristine gas, unpolluted by 
metals.  

 The above observational evidence thus strongly suggests that there exists in 
our local volume of space a few young dwarf galaxies in which the first star 
formation did not occur until $\sim$ 100 Myr ago. Somehow the H I clouds with 
which they are associated have not undergone gravitational collapse for a whole 
Hubble time. The detailed studies of these young dwarf galaxies are not only 
important for understanding their intrinsic properties,
but are also crucial for galaxy formation studies. The proximity of these young 
dwarfs allows studies of their structure, metal content and stellar 
populations, and star formation processes in a nearly pristine environment with 
a sensitivity, precision and spatial resolution that faint small angular size 
distant high-redshift galaxies do not allow.
We focus here on the analysis of heavy element abundance ratios in BCGs. We 
shall show that some of these ratios are very good indicators of the 
BCG age.

   It is now well established that the 
oxygen seen in H II regions is a primary element produced by massive stars
with $M$ $\geq$ 10 $M_\odot$ (see the review by Weaver \& Woosley 1993). As for
the other $\alpha$-process elements seen in the spectra of H II regions, 
such as neon, silicon, sulfur and argon, they are generally thought to be 
also primary (see Thuan, Izotov \& Lipovetsky (1995, hereafter TIL95) for 
references).   
Carbon is also a primary element, believed to be produced by both massive and 
intermediate-mass (3 $M_\odot$ $\leq$ $M$ $\leq$  9 $M_\odot$)
stars. However, because carbon has its strong nebular emission lines
not in the optical but the UV range, it has not been possible until recently, 
with the advent of {\sl HST}, to obtain precise measurements
of its abundance. Garnett
et al. (1995a) found the C/O ratio in BCGs to increase with increasing oxygen 
abundance and its value in the lowest metallicity 
galaxies to be below the one predicted by stellar
nucleosynthesis theory. However, a subsequent study by Garnett et al. (1997)  
found the C/O ratio in I Zw 18 to be  
much larger (by a factor of $\sim$ 2) than the mean value for other
metal-deficient galaxies. This led those authors to conclude that I Zw 18 is not 
a young galaxy, but that it has experienced carbon-enriching episodes of star 
formation in the past.

   The situation for nitrogen is even more complex. 
Spectral observations of
H II regions in spiral galaxies with high metallicities 
($Z$ $\sim$ $Z_\odot$)
show that the N/O ratio increases with O/H ( Pagel \& Edmunds 1981; Serrano
\& Peimbert 1983; Torres-Peimbert, Peimbert \& Fierro 1989; Vila-Costas \&
Edmunds 1993). This implies that a large part of the nitrogen in these
high-metallicity spiral galaxies is produced as a secondary element in the
CNO cycle, i.e. its synthesis is controlled by carbon seed atoms already present
in the main-sequence or red-giant stars (which can be of any mass). However,
in low metallicity ($Z_\odot$/50 $\leq$ $Z$ $\leq$ $Z_\odot$/3) 
environments such as in irregular and blue compact galaxies, the N/O ratio is 
found to be constant and independent of O/H ( Lequeux et al. 1979; French 1980; 
Kinman \& Davidson 1981; Kunth \& Sargent 1983; Campbell, Terlevich \& Melnick 
1986; TIL95). This implies that nitrogen is mainly a primary element 
when O/H is small. Stellar nucleosynthesis models predict that 
primary nitrogen is produced mainly in intermediate-mass stars by hot-bottom 
carbon burning undergoing a third dredge-up episode which brings 
carbon-rich material from the core into the hydrogen-burning shell 
(Renzini \& Voli 1981, hereafter RV81; van den Hoek \& Groenewegen 1997,
hereafter HG97). The spread of the
N/O ratio at a fixed O/H in early spectral observations
is seen to be large and has been attributed to
observational uncertainties and/or to the time delays
($\tau$ $\leq$ 5$\times$10$^8$ yr) between the production of oxygen by 
short-lived massive stars and that of nitrogen by longer-lived 
intermediate-mass stars ( Matteucci \& Tosi 1985; 
Garnett 1990; Pilyugin 1992, 
1993; Vila-Costas \& Edmunds 1993; Marconi, Matteucci 
\& Tosi 1994). However, TIL95 have shown subsequently that a large part of the 
scatter is due to observational uncertainties. High signal-to-noise
observations show a very small dispersion of log N/O ($\pm$ 0.02 dex) at low 
metallicities, which favors primary nitrogen production not in 
intermediate-mass stars but in massive ones.

   The optical spectra of BCGs also contain iron lines, which allow to derive
abundances for that element, as was first done by TIL95. Iron is a primary 
element produced by explosive nucleosynthesis either in type I supernovae (SNe 
I) with low-mass progenitors, or in type II supernovae (SNe II) with more 
massive progenitors (Weaver \& Woosley 1993; Woosley \& Weaver 1995,
hereafter WW95).
During the collapse of the Galaxy, while O and other $\alpha$-elements are
mainly produced in massive stars and restored to the interstellar 
medium (ISM) through SNe II, iron is thought to come mainly from SNe I. 
Therefore, due to the different evolution timescales of stars with different 
masses, oxygen is injected sooner than iron into the ISM, resulting in an oxygen 
overabundance relative to iron in the halo population. It is well established 
that halo Population II stars with [Fe/H] $<$ --1 ( where [X] $\equiv$ log X -- 
log 
X$_\odot$) have a constant overabundance [O/Fe] $\sim$ 0.3 -- 0.5. As for disk 
stars with higher metallicities ([Fe/H] $\geq$ --1), they show a progressive 
decrease of O/Fe toward the solar ratio ([O/Fe] $\sim$ 0) (Lambert, Sneden \& 
Ries 1974; Spite \& Spite 1981; Gratton \& Ortolani 1986; Barbuy 1988; 
Barbuy \& Erdelyi-Mendes 1989; Edvardsson et al. 1993). 

   In order to constrain models of stellar nucleosynthesis and chemical
evolution of the heavy elements discussed above in a metal-deficient 
environment, we present in this paper high signal-to-noise ratio
spectrophotometric observations of 54 H II regions in 50 low-metallicity 
($Z_\odot$/50 $\leq$ $Z$ $\leq$ $Z_\odot$/7) BCGs. This paper is a continuation
of the study by TIL95, based on a substantially larger observational sample 
(more than triple the size of the old sample). We found the enlarged sample to 
confirm and considerably strengthen the main results obtained by TIL95. In \S2 
we discuss the properties of the sample. The methods used for heavy
element abundance determinations are discussed in \S3. In \S4 we discuss how the 
behavior of various heavy element abundance ratios as a function of metallicity 
can constrain the origin of these elements. We discuss in \S5 how these data can 
put limits on the age of BCGs. We show that BCGs with metallicities below a 
certain value are likely to be young systems which are making stars for the 
first time at the present epoch.  
In \S6 we compare our results for heavy element abundances in BCGs with those in 
another highly metal-deficient environment: that of high-redshift damped 
Ly$\alpha$ galaxies. We summarize our results in \S7.

\section {THE SAMPLE}

   We choose for analysis the sample of low-metallicity 
BCGs for which we have obtained optical spectrophotometry to derive the 
primordial helium abundance. Because the requirements for an accurate helium 
abundance determination are very strict (the precision has to be better than a 
few percent), the spectra have been obtained at the 4m and 2.1m Kitt Peak 
telescopes with the very highest signal-to-noise ratio, typically $\sim$ 20 to 
40 in the  continuum. 
The sample includes mainly BCGs from the First Byurakan (or Markarian) and 
Second Byurakan surveys. Additionally, UGC 4483, VII Zw 403 and 
several BCGs from the University of Michigan (UM) survey have been included. All 
galaxies have been observed
with nearly the same observational settings (slit width $\sim$ 2\arcsec,
spectral resolution of 6 \AA, and spectral range covered during a single exposure 
$\lambda$$\lambda$3600 -- 7500 \AA), and the data reduction has been done
in a homogeneous way. The results of the spectrophotometric observations and the 
 details of the data reduction are presented in Izotov, Thuan \& Lipovetsky 
(1994, 1997, hereafter ITL94 and ITL97), TIL95 and
Izotov \& Thuan (1998ab). The total sample consists of 54 supergiant H II
regions observed in 50 BCGs. All observed H II 
regions are of high excitation, 
with equivalent widths of H$\beta$ generally larger than 100 \AA.
For the majority of the BCGs in the sample, high-order H9 to H12 Balmer emission
lines are detected, which permits to correct the observed line intensities  
for both extinction and underlying hydrogen stellar absorption with a very high
precision over the whole optical spectral range. This is most important as 
uncertainties in these corrections will introduce uncertainties in the derived 
abundances, resulting in an artificial scatter in the data. 
In particular, it is especially important that these corrections are done 
correctly for the [O II] $\lambda$3727 line at the blue end of the spectrum. 
This line is used for the determination of the O$^+$ abundance
and for the correction of nitrogen and iron abundances for unseen stages
of ionization. Ignoring, for example, stellar hydrogen absorption will lead to 
an overestimate of the extinction and of the intensity of
[O II] $\lambda$3727, and to an underestimate of nitrogen and iron abundances.
In all cases therefore, we have made sure that the Balmer hydrogen emission line 
intensity ratios are in agreement with the theoretical recombination ratios 
after correction for both these effects. 

    Another important feature of the sample is that the 
auroral [O III] $\lambda$4363 line is detected with a high signal-to-noise 
ratio. This 
allows to derive electron temperatures and element 
abundances with a much higher precision than with the use of empirical 
statistical methods based, for instance, on
the total intensity of oxygen lines. The precision of these empirical methods is 
not better than 0.2 -- 0.3 dex in the determination of the oxygen abundance.

    Our sample includes nearly all of the most metal-deficient BCGs known. Among 
the catalogued BCGs with heavy element abundances less than 1/20 solar, only 3 
galaxies (CG 1116+51, Tol 65 and Tol 1214--277) are missing. The other 6 are 
present, including the two most metal-deficient galaxies known, I Zw 18 and SBS 
0335--052. We have purposely not included galaxies observed by other authors in 
the optical range so as to preserve the homogeneity of a data set obtained and 
reduced by us in a uniform way. This allows us to minimize the scatter which 
would be inevitably and artificially introduced, were we to assemble data by 
different authors with varying quality, which differ by their reduction methods,  
their line intensity measurements and corrections for extinction and underlying 
stellar absorption, the formulae used for abundances calculations, etc.

    For the derivation of carbon and silicon abundances, we have assembled a 
sample of BCGs observed in the UV with the {\sl HST} Faint Object Spectrograph
(FOS) by Garnett et al. (1995ab, 1997), Kobulnicky et al. (1997),
Kobulnicky \& Skillman (1998) and Thuan, Izotov
\& Foltz (1998). In total, this sample includes 15 supergiant H II regions
in 11 galaxies. Again, in an effort to treat the data in a homogeneous way, we 
have reanalyzed all the UV data in the same manner, supplementing it for some 
galaxies with our most recent ground-based observations.

\section{HEAVY ELEMENT ABUNDANCE DETERMINATION}

\subsection{Ground-based optical observations}

To derive element abundances from the optical spectra, we have followed the 
procedure detailed in ITL94 and ITL97. We adopt a two-zone photoionized H II 
region model (Stasi\'nska 1990): a high-ionization zone with temperature $T_e$(O 
III), and a low-ionization zone with temperature $T_e$(O II).  We have  
determined $T_e$(O III) from the 
[O III]$\lambda$4363/($\lambda$4959+$\lambda$5007) ratio 
using a five-level atom model (Aller 1984)
with atomic data from Mendoza (1983). That temperature is used for the 
derivation of the He$^+$, He$^{2+}$, O$^{2+}$, Ne$^{2+}$ and Ar$^{3+}$ ionic 
abundances. To derive $T_e$(O II), we have used 
the relation between
$T_e$(O II) and $T_e$(O III) (ITL94), based on a fit to the
photoionization models of Stasi\'nska (1990). The temperature $T_e$(O II) is 
used to derive the O$^+$, N$^+$ and Fe$^+$ ionic abundances. For Ar$^{2+}$
and S$^{2+}$ we have used an electron temperature intermediate between
$T_e$(O III) and $T_e$(O II) following the prescriptions of Garnett (1992).
The electron density $N_e$(S II) is derived from
the [S II] $\lambda$6717/$\lambda$6731 ratio.

Total element abundances are derived after correction for unseen stages
of ionization as described in ITL94 and TIL95.
In the spectra of several BCGs, a strong nebular He II $\lambda$4686 emission
line is present, implying the presence of a non-negligible amount of O$^{3+}$. 
Its abundance is derived from the relation:
\begin{equation}
{\rm O}^{3+}=\frac{{\rm He}^{2+}}{{\rm He}^+}({\rm O}^++{\rm O}^{2+}). 
\label{eq:O3+}
\end{equation}
Then, the total oxygen abundance is equal to
\begin{equation}
{\rm O}={\rm O}^++{\rm O}^{2+}+{\rm O}^{3+}.       \label{eq:O}
\end{equation}
The electron temperatures, number densities, ionic and total element
abundances along with ionization correction factors for each
galaxy in our sample are given in ITL94, TIL95, ITL97, and Izotov \&
Thuan (1998b). Because the electron temperatures $T_e$(O III) in ITL94, TIL95 
and ITL97 are calculated using a three-level atom model, while Izotov \&
Thuan (1998b) adopt a five-level atom model (Aller 1984), we have recalculated, for 
consistency, all heavy element abundances for the galaxies in ITL94, TIL95 and 
ITL97 using also a five-level atom model. This gives slightly
higher electron temperatures and slightly lower oxygen abundances (by $\sim$ 0.04 
dex). The heavy element abundance ratios are however nearly unchanged.
Table 1 lists the oxygen abundance and the 
heavy element to oxygen abundance ratios for all the galaxies in our sample. 

\subsection{{\sl HST} UV and optical observations}

    Carbon is one of the elements which provides strong constraints on the 
chemical evolution of BCGs and the origin of elements. However, it possesses no 
strong optical emission line. The strongest carbon emission line is the 
C III] $\lambda$(1906+1909) line in the UV. With the advent of {\sl HST}, much 
work has been done to determine the carbon abundance in BCGs, using 
FOS spectroscopy (Garnett et al. 1995a, 1997; Kobulnicky
et al. 1997 and Kobulnicky \& Skillman 1998). The same spectra can also be used 
to derive the silicon abundance when the Si III] 
$\lambda$(1885+1892) line is seen. 
We have decided to reanalyze some of these data, specifically the galaxies with 
both UV and optical FOS spectra, for the following reasons:  
 (1) Garnett et al. (1995a) used the UV O III] $\lambda$1666
to derive the C/O abundance ratio. While this method has the advantage of being 
insensitive to the extinction curve adopted, the O III] $\lambda$1666 and C III] 
$\lambda$1906+1909 emission lines being close to each other in wavelength,
the weakness of the O III] $\lambda$1666 line makes in many cases the C/O 
abundance ratio determination very uncertain.
Therefore, we have used the [O III]
$\lambda$(4959 + 5007) emission lines in the FOS spectra (these are used instead 
of higher S/N ground-based spectra to insure that exactly the same region is 
observed as in the UV) to determine the C/O abundance ratio. This approach is 
subject to uncertainties in the extinction curve but, fortunately, at least in 
some of the galaxies, the extinction is small.
(2) Because of the uncertainties in the atomic data, different authors have used 
different analytical expressions for the determination of carbon and silicon
abundances. As 
this may introduce systematic shifts, we have reanalyzed all spectra in the same 
manner, using the same expressions, to insure the homogeneity of our data set.  
(3) The C/O abundance ratio is dependent on several physical parameters, and in 
particular on the electron temperature which needs to be determined with a 
relatively high precision. Because of the poorer signal-to-noise ratio of the 
FOS spectra in the optical range, Garnett et al. (1997) and Kobulnicky \& 
Skillman (1998) have used electron temperatures derived from higher 
signal-to-noise ratio ground-based observations. We use here our own  
new ground-based observations with very high signal-to-noise ratio to better 
constrain the electron temperature for some of the most metal-deficient galaxies
in our sample.
(4) In deriving silicon abundances, Kobulnicky et al. (1997) have not corrected  
for unseen stages of ionization. We have taken into account here the ionization 
correction factor for Si which can be large.

We derive the C$^{2+}$ abundance from the relation (Aller 1984):
\begin{equation}
\frac{{\rm C}^{2+}}{{\rm O}^{2+}}=0.093\exp\left(\frac{4.656}{t}\right)
\frac{I({\rm C III]}\lambda1906+\lambda1909)}
{I({\rm [O III]}\lambda4959+\lambda5007)},
\label{eq:C2+}
\end{equation}
where $t$ = $T_e$/10$^4$. Following Garnett et al. (1995a) we adopt
for the temperature in Eq. (\ref{eq:C2+}) the $T_e$(O III) value in
the O$^{++}$ zone. Then
\begin{equation}
\frac{\rm C}{\rm O}={\rm ICF}\left(\frac{\rm C}{\rm O}\right)
\frac{{\rm C}^{2+}}{{\rm O}^{2+}}. 
\label{eq:CO}
\end{equation}
The correction factor ICF(C/O) in Eq.(\ref{eq:CO}) for unseen ionization 
stages of carbon is taken from Garnett et al. (1995a). 
In the majority of cases, the correction factor is 
small, i.e. 
ICF(C/O) = 1.0 -- 1.1. However, it is evident from Eq. (\ref{eq:C2+}) that 
the C/O abundance ratio is critically dependent on the electron temperature. To 
derive as precise temperatures as possible, we have used, when available, very 
high signal-to-noise ground-based observations obtained within apertures nearly 
matching the circular or square 0\farcs86 FOS aperture.

The abundance of silicon is derived following Garnett et al. (1995b) from the
relation
\begin{equation}
\frac{\rm Si}{\rm C}={\rm ICF}\left(\frac{\rm Si}{\rm C}\right)
\frac{{\rm Si}^{2+}}{{\rm C}^{2+}},
\label{eq:Si}
\end{equation}
where
\begin{equation}
\frac{{\rm Si}^{2+}}{{\rm C}^{2+}}=0.188t^{0.2}\exp\left(\frac{0.08}{t}\right)
\frac{I({\rm Si III]}\lambda1883+\lambda1892)}
{I({\rm C III]}\lambda1906+\lambda1909)}.
\label{eq:Si2+}
\end{equation}
For the determination of the silicon abundance, we again adopt the temperature 
$T_e$(O III). The correction factor ICF(Si/C) is given by
Garnett et al. (1995b) and is larger than that for carbon. 
One of the main uncertainties in the determination of the silicon abundance 
comes from the Si III] $\lambda$1892 emission line. It is not seen in some
galaxies (such as in the NW component of I Zw 18), while atomic theory predicts  
that this line should be present, given the signal-to-noise ratio and intensity 
of the neighboring Si III] $\lambda$1883 emission line.
In those cases, we assume
$I$(Si III] $\lambda$1883 + $\lambda$1892) = 
1.67 $\times$ $I$(Si III] $\lambda$1883) as expected in the low-density limit.

We discuss next in more details the carbon and silicon abundance determinations
in a few BCGs of special interest. The heavy element abundance ratios derived  
from {\sl HST} FOS observations are given in Table 4.

\subsubsection{I Zw 18}

   Both the NW and SE components of this BCG, the most metal-deficient galaxy
known, have been observed in the UV and optical with the {\sl HST} FOS by 
Garnett et al. (1997). Adopting $T_e$(O III) = 19,600 K and 17,200 K  respectively 
for the NW and SE components (Skillman \& Kennicutt 1993) and using the C III] 
$\lambda$(1906+1909) and [O III] $\lambda$(4959+5007) emission lines, those 
authors derived very high C/O abundance ratios for both components,
--0.63$\pm$0.10 and --0.56$\pm$0.09, as compared to $\sim$ --0.8 in other
metal-deficient galaxies and predictions of massive stellar 
nucleosynthesis theory (Weaver \& Woosley 1993; WW95).
This led Garnett et al. (1997) to conclude that I Zw 18 is
not a young galaxy and has experienced star formation episodes in the past which 
have enhanced the C/O ratio through the evolution of intermediate-mass
stars.

   As the adopted value for the electron temperature is crucial for the 
determination of the C/O abundance ratio, we have redetermined it 
using new data by Izotov et al. (1997b). Those authors have obtained with the 
Multiple Mirror Telescope (MMT) a spectrum of both NW and SE components during 
a 3-hour exposure with excellent seeing conditions (FWHM $\sim$ 0.7 arcsec). 
The much higher signal-to-noise ratio of this spectrum as compared to that by 
Skillman \& Kennicutt (1993) has resulted in the discovery of a Wolf-Rayet 
stellar population in the NW component of I Zw 18 (Izotov et al. 1997b). We have 
extracted from this two-dimensional spectrum two one-dimensional spectra at the 
location of the brightest parts of the NW and SE components, within the smallest 
aperture allowed by the MMT observations, that of 0\farcs6$\times$1\farcs5. This 
provides a fairly good match to the round 0\farcs86 FOS aperture, as the ratio 
of the area of the former to that of the latter is $\sim$ 1.54.  
The observed and corrected emission line fluxes relative to H$\beta$ for both 
components are shown in Table 2 together with the extinction coefficients 
$C$(H$\beta$), observed H$\beta$ fluxes and equivalent widths, and equivalent 
widths of the underlying stellar Balmer absorption lines. 
The errors of the line intensities listed in Table 2 take into account the noise 
statistics in the continuum, the errors in placing the continuum and fitting the 
line profiles with gaussians. 
We also retrieved  
the FOS spectra from the {\sl HST} archives and remeasured the 
emission line fluxes.  Comparison of the MMT and FOS H$\beta$ fluxes in the SE 
component shows very good agreement: 3.5$\times$10$^{-15}$ ergs 
cm$^{-2}$s$^{-1}$ in the MMT spectrum as compared to 3.2$\times$10$^{-15}$ ergs 
cm$^{-2}$s$^{-1}$ in the {\sl HST} spectrum. The H$\beta$ emission equivalent 
widths are also in good agreement: 144 \AA\ in the MMT spectrum as compared to 
127 \AA\ in the {\sl HST} spectrum.
There is not so good agreement, however, for the NW component. The ratio of the 
observed H$\beta$ flux in the MMT spectrum of 2.9$\times$10$^{-15}$ ergs 
cm$^{-2}$s$^{-1}$ to that of 5.0$\times$10$^{-15}$ ergs 
cm$^{-2}$s$^{-1}$ in the {\sl HST} spectrum is $\sim$ 0.6, just the ratio of the 
measured emission equivalent width of H$\beta$ of 34 \AA\ from
the MMT spectrum to that of 55 \AA\ from the {\sl HST} spectrum.
These differences are probably due to a slight positioning shift between the 
MMT and {\sl HST} apertures on the NW component.
  The electron temperatures and oxygen abundances derived from the MMT spectra 
for both NW and SE components are shown in Table 3. 
In deriving these quantities, we have neglected temperature fluctuations. While 
there is evidence for large temperature fluctuations in planetary nebulae and 
perhaps in high-metallicity HII regions, we believed that there is, until now,
no such convincing evidence for low-metallicity HII regions like the ones 
considered here (see a detailed discussion of this issue in Izotov \& Thuan 1998b). 
The electron temperatures
$T_e$(O III) are much higher than those obtained by Skillman \& Kennicutt (1993) 
through a larger aperture (2\arcsec$\times$5\arcsec\ for the SE component and 
2\arcsec$\times$7\farcs55 for the NW component), by 1900 K in the NW component, 
and by 2300 K in the SE component. They are 
also higher by 1800 K and 700 K respectively than the temperatures obtained by 
Izotov \& Thuan (1998a) within an aperture of 2\arcsec$\times$5\arcsec. 
There is evidently a temperature gradient at the center of both NW and SE 
components (see also Martin 1996), 
and large apertures give invariably lower electron temperatures. 
Matching the small {\sl HST} FOS aperture as closely as possible is essential to 
derive appropriate abundances. The higher electron temperatures lead to 
lower oxygen abundances than those derived by Skillman \& Kennicutt (1993) and 
Izotov \& Thuan (1998a) for larger regions. Note that the oxygen abundance 
derived for the NW component is lower than that derived for the SE component by 
a factor of $\sim$ 1.2. This is again because there is a gradient 
toward higher temperatures in the central part of the NW component, which leads 
to lower abundances. In larger apertures, the derived oxygen abundances are the 
same within the errors for both components, suggesting good mixing (Skillman \& 
Kennicutt 1993; Izotov \& Thuan 1998a). The new C/O abundance ratio derived 
using emission line fluxes from Garnett et al. (1997) is also lower than the 
value derived by those authors, implying 
C/O abundance gradients. 
We have also measured the fluxes of the Si III] $\lambda$1883, 1992 emission
lines and derived silicon abundance in both components.

\subsubsection{SBS 0335--052}

    Because Garnett et al. (1995a) observed SBS 0335--052 only
in the UV, they used the C III] $\lambda$(1906 + 1909)
to O III] $\lambda$1666 flux ratio to derive log C/O =
--0.94$\pm$0.17, the lowest value for the BCGs in 
their sample. Although the O III] $\lambda$1666 was clearly
detected, the UV spectrum is noisy, especially in the blue part, introducing 
uncertainties on the line strength. Garnett et al. (1995b) have also 
derived a low Si/O abundance: log Si/O = --1.72$\pm$0.20. This low abundance 
ratio is mainly due to the low value of C/O, as Si/O is derived using the Si/C 
ratio.

    We again use new high signal-to-noise ratio MMT spectral observations  
(Izotov et al. 1997a) in an effort to improve the situation. We extracted
a one-dimensional spectrum of the brightest part of the SBS 0335--052 within 
a 1\arcsec$\times$0\farcs6 aperture to best match the FOS aperture. We use the 
MMT spectrum to derive the electron temperature and the oxygen abundance, and 
combine with the the {\sl HST} UV spectrum to derive C/O and Si/O abundance 
ratios. Emission line fluxes are shown in Table 2, while the 
electron temperature $T_e$(O III) and oxygen abundance are given in 
Table 3. The observed C III] $\lambda$(1906+1909) and Si III] 
$\lambda$(1882+1893) emission line fluxes are corrected for extinction using the 
 reddening law for the Small Magellanic Cloud, as parameterized by
Pr\'evot et al. (1984). The carbon and silicon abundances are shown in Table 4.
We derive higher log C/O = --0.83$\pm$0.08 and log Si/O = --1.60$\pm$0.21
values as compared to those of Garnett et al. (1995ab), although they are 
consistent within the errors.

\subsubsection{SBS 1415+437}

    This galaxy, with a heavy element abundance $Z_\odot$/21 (Izotov \&
Thuan 1998b), has been observed 
with the {\sl HST} FOS in the UV and optical ranges (Thuan, Izotov \& Foltz 
1998). As in I Zw 18 NW, the Si III] $\lambda$1892 emission line is not seen. 
Instead, at the location of this line a deep absorption is observed. Therefore, 
the total flux of the Si III] emission lines has been derived by multiplying the 
Si III] $\lambda$1883 emission line flux by a factor of 1.67. The 
UV emission line fluxes have been corrected for extinction using
the reddening law for the Small Magellanic Cloud (Pr\'evot et al. 1984). The 
electron temperature is derived from a high signal-to-noise ratio MMT spectrum 
in a 1\farcs5$\times$0\farcs6 aperture (Thuan, Izotov, \& Foltz 1998). 

\subsubsection{UM 469, NGC 4861 and T1345--420}

    We use for these galaxies the corrected emission line fluxes derived
by Kobulnicky \& Skillman (1998) to calculate the C/O and Si/O abundance
ratios with equations (\ref{eq:C2+} -- \ref{eq:Si2+}). We also 
correct the C$^{++}$/O$^{++}$ abundance ratio for unseen stages of
ionization while Kobulnicky \& Skillman (1998) decided not to apply a correction 
factor.

\subsubsection{NGC 5253}

    Kobulnicky et al. (1997) have derived C/O and Si/O abundance ratios in three 
H II regions of this galaxy. Our measurements
of the emission line fluxes in the same FOS spectra retrieved from the {\sl HST} 
archives are in fair agreement with theirs, except for
the Si III] $\lambda$(1883+1892) emission line flux in the HII-2 region, 
for which we derive a higher value. We have also corrected
Si$^{++}$/C$^{++}$ for unseen stages of ionization, which was not done by 
Kobulnicky et al. (1997). This correction factor can be as high as $\sim$ 1.4.

\section{HEAVY ELEMENT ABUNDANCES}

    Our main goal here is to use the large homogeneous BCG sample described 
above to extend the work of TIL95 and study in more detail and with more 
statistics the relationship between different heavy elements in a very 
low-metallicity environment. Some of the very low-metallicity galaxies are most 
likely young nearby dwarf galaxies undergoing their first burst of star 
formation not more than 100 Myr ago (Thuan, Izotov
\& Lipovetsky 1997; Thuan \& Izotov 1997; Thuan, Izotov \& Foltz 1998).
Therefore, the relationships between different heavy element abundances will not 
only put constraints on the star-formation history of BCGs, they will also be 
useful for understanding the early chemical evolution of galaxies. Furthermore, 
a precise determination of heavy element abundances in BCGs can put constraints
on stellar nucleosynthesis models, as theoretical predictions for the yields
of some elements are not yet very firm. 
 The best studied and most easily observed element in BCGs is oxygen. 
Nucleosynthesis theory predicts it to be produced only by massive stars. We 
shall use it as the reference chemical element and consider the behavior of 
heavy element abundance ratios as a function of oxygen abundance.

\subsection{Neon, Silicon, Sulfur and Argon}

    The elements neon, silicon, sulfur and argon are all products of
$\alpha$-processes during both hydrostatic and explosive nucleosynthesis
in the same massive stars which make oxygen. Therefore, the Ne/O, Si/O,
S/O and Ar/O ratios should be constant and show no dependence on the oxygen 
abundance. In Figure 1 we show 
the abundance ratios for these elements as a function of 12 + log O/H 
(filled circles). For Si/O, we have also shown for comparison the data from 
Garnett et al. (1995b) for those galaxies which we have not
reanalyzed for lack of the necessary data (open circles). These galaxies are not 
included in the computation of 
the mean abundance ratios listed in Table 5. 
The mean values of these element abundance ratios are directly related to the 
stellar yields and thus provide strong constraints on the theory of massive 
stellar nucleosynthesis. Note, that while silicon and sulfur abundances can be 
measured in a wide variety of astrophysical settings (stars, H II regions,
high-redshift damped Ly$\alpha$ clouds), neon and argon abundances can be 
measured with good precision at low metallicities only in BCGs. Table 5 gives 
the mean values of the Ne/O, Si/O, S/O and Ar/O abundance ratios calculated for 
the total sample, as well as for two subsamples: a low-metallicity subsample 
containing galaxies with 12 + log O/H $\leq$ 7.6, and a high-metallicity 
subsample containing galaxies with 12 + log O/H $>$ 7.6. For comparison, we list 
also the solar ratios taken from Anders \& Grevesse (1989). Generally, the 
dispersion about the mean of the points in the low-metallicity 
subsample is smaller than that in 
the high-metallicity subsample and the total sample, although 
the differences are not statistically significant.
Examination of Figure 1 and Table 5 shows that, as predicted by stellar 
nucleosynthesis theory, no dependence on oxygen abundance is found for any
of the Ne/O, Si/O, S/O and Ar/O abundance ratios.

 Independently
of the subsample, the Ne/O, Si/O, S/O and Ar/O ratios are all  
very close to the corresponding solar value (dotted lines in Figure 1). 
There may be a hint that the mean 
Si/O abundance ratio in BCGs is slightly lower than the solar value, but the  
difference is again not statistically significant. We thus conclude that there 
is no significant depletion of silicon into dust grains in BCGs. 
By contrast, Garnett et 
al. (1995b) have found a weighted mean log Si/O = --1.59$\pm$0.07 for their 
sample of low-metallicity galaxies, lower by a factor of $\sim$ 1.6 than the 
solar value. This led them to conclude that about 50\% of the silicon is 
incorporated into dust grains. The number of galaxies with
measured silicon abundances is not large, and more observations are needed for a 
more definite conclusion.

The mean values of the abundance ratios for the other elements are in very good
agreement with those derived by TIL95 for a smaller sample of BCGs. TIL95 have 
made detailed comparisons of their results with those of previous studies,
so we shall not repeat the discussion here.
We shall only mention the more recent work of van Zee, Haynes \& Salzer (1997) 
who 
have studied heavy element abundance ratios in the H II regions of 28 gas-rich, 
quiescent dwarf galaxies. Their mean Ne/O, S/O and Ar/O ratios are consistent 
within the errors with the results of TIL95 and those obtained here, although  
in many of their galaxies, the abundance measurements are more uncertain because 
the [O III] $\lambda$4363 emission line is not detected and the electron 
temperature is derived from an empirical method.

\subsection{Carbon}

   Carbon is produced by both intermediate and high-mass stars. 
Since C is a product of hydrostatic burning, the 
contributions of SNe Ia and SNe II are small. Therefore, the C/O abundance ratio 
is sensitive to the particular star formation history of the galaxy.
It is expected that, in the earliest stages of galaxy evolution, carbon is 
mainly produced by massive stars, so that  
the C/O abundance ratio is independent of the oxygen
abundance, as both C and O are primary elements. At later stages, 
intermediate-mass stars add their carbon production, so that an 
increase in the C/O ratio is expected with increasing oxygen abundance.

  The results of the study by Garnett et al. (1995a) did not conform to these 
expectations. Those authors found a continuous increase
of log C/O with increasing log O/H in their sample of metal-deficient galaxies, 
which could be fitted by a power law with slope
0.43. Because log C/O is fairly constant at --0.9 for halo stars in the 
Galaxy (Tomkin et al. 1992), Garnett et al. (1995a) suggested there may be a 
difference between the abundance patterns seen in the Galaxy and dwarf galaxies.
Subsequent {\sl HST} FOS observations of I Zw 18 by Garnett et al. (1997) 
yielded abundances that  
complicated the situation even more. It was found that I Zw 18 bucks the trend 
shown by the other low-metallicity objects. Although it has the lowest 
metallicity known, it shows a rather high log C/O, equal to --0.63$\pm$0.10 and 
--0.56$\pm$0.09 in the NW and SE components respectively. These values are 
significantly higher than those predicted by massive stellar nucleosynthesis 
theory. This led Garnett et al. (1997) to conclude that carbon in I Zw 18 has 
been enhanced by an earlier population of lower-mass stars and, hence, despite 
its very low metallicity, I Zw 18 is not a ``primeval" galaxy. Garnett et al. 
(1997) considered and dismissed the possibility that the high C/O ratio in I Zw 
18 may be due to errors in the electron temperature. We have revisited the 
problem here for I Zw 18 with our own data and conclude that it is indeed a too 
low adopted electron temperature which is responsible for the too high C/O ratio 
in both components of I Zw 18. The temperatures adopted by Garnett et al. (1997) 
for I Zw 18 are derived from optical observations in an aperture larger than the 
FOS aperture, and are too low because of a temperature gradient.
 In Figure 2a
we have plotted with filled circles log C/O against 12 + log O/H for all the 
galaxies which we have reanalyzed. Great care was taken to derive electron 
temperatures in apertures matching as closely as possible the FOS aperture. Open 
circles show 
galaxies from Garnett et al. which could not be reanalyzed
because of lack of the necessary data. It is clear that, in contrast to the 
results by Garnett et al. (1995a, 1997), we find log C/O to be constant in the  
extremely low-metallicity range, when 12 + log O/H varies between 7.1 and 7.6, 
as expected from the common origin of carbon and oxygen in massive stars. 
Furthermore, the dispersion of the points about the mean is very small : $<$log 
C/O$>$ = --0.78$\pm$0.03 (Table 5). This mean value is 
in very good agreement with 
that of $\sim$ --0.8 predicted by massive stellar nucleosynthesis theory (Weaver 
\& 
Woosley 1993; WW95). Two models with $Z$ = 0 and $Z$ = 0.01 
$Z_\odot$ are shown by horizontal lines in Figure 2a. They are in good agreement 
with the observations.   
 At higher metallicities (12 + log O/H $>$ 7.6), there is an increase in log 
C/O with log O/H and also more scatter at a given O/H, which we attribute to the 
carbon contribution of intermediate-mass stars in addition to that of massive 
stars.

\subsection{Nitrogen}

    The origin of nitrogen has been a subject of debate for some years. The 
basic nucleosynthesis process is well understood -- nitrogen
results from CNO processing of oxygen and carbon during hydrogen burning --
however the nature of the stars mainly responsible for the production of 
nitrogen remains uncertain. If oxygen and carbon are produced not in previous 
generation stars, but 
in the same stars prior to the CNO cycle, then the amount of nitrogen produced 
is independent of the initial heavy element abundance of the
star, and its synthesis is said to be primary. On the other hand, if the "seed" 
oxygen and carbon are produced in previous generation stars and incorporated 
into a star at its formation and a constant
mass fraction is processed, then the amount of nitrogen produced is
proportional to the initial heavy element abundance, and the nitrogen
synthesis is said to be secondary. Secondary nitrogen synthesis can occur
in stars of all masses, while primary nitrogen synthesis is usually
(but not universally)
thought to occur mainly in intermediate-mass stars (RV81, WW95). 
In the case of secondary 
nitrogen production, it is expected that massive
stars with decreasing metallicity would produce decreasing
amounts of $^{14}$N (WW95). There is, however, a caveat: there 
is the possibility that in some massive stars the convective helium shell 
penetrates into the hydrogen layer, with the consequent production of large 
amounts of primary nitrogen.

   The behavior of the N/O abundance ratio as a function of the O/H ratio has 
provided the main observational constraint to this debate. In low-metallicity 
Galactic halo stars, N/O is nearly independent of O/H implying that nitrogen has 
a strong primary component which can be
explained by primary nitrogen production in massive stars with
a large amount of convective overshoot (Timmes, Woosley \& Weaver 1995).
Several studies of low-metallicity (12 + log O/H $\leq$ 8.3) H II regions in 
dwarf galaxies have also revealed that N/O is independent on O/H implying again 
a primary origin of nitrogen (Garnett 1990; Vila-Costas  \& Edmunds 1993; TIL95; 
van Zee, Haynes \& Salzer 1997; van Zee, Salzer \& Haynes 
1998). It is generally believed that primary nitrogen in BCGs is
produced by intermediate-mass  
stars by carbon dredge-up (RV81). For high-metallicity H II 
regions in spiral galaxies (12 + log O/H $>$ 8.3), the N/O ratio increases 
linearly with the O abundance, indicating that, in this metallicity regime, N is 
primarily a secondary element. We shall not be concerned with secondary N here 
and shall discuss mainly primary N since all our BCGs have 12 + log O/H $<$ 8.3.

 A problem in interpreting the data in the vast majority of 
studies is the existence  
of a considerable scatter ($\pm$ 0.3 dex) in N/O at a given O/H.
The large scatter has been attributed to the delayed release of N produced in 
intermediate-mass long-lived stars, compared to O produced in massive 
short-lived stars (Matteucci \& Tosi 1985; Matteucci 1986; Garnett 1990; 
Pilyugin 1992, 1993; Marconi et al. 1994; Kobulnicky \& Skillman 1996, 1998). 
In contrast to these studies, TIL95 found the scatter in N/O at a given O/H 
to be large only when the metallicity exceeds a certain value, i.e. 12 + log O/H 
$>$ 7.6. At lower O abundances, the scatter is extremely small. It is 
difficult to improve much the statistics by adding to the TIL95 sample more 
galaxies in this very low-metallicity range because the known objects having 
such a low heavy element content are very scarce and extremely difficult to 
discover.  We have added only SBS 
0335--052 with $Z_\odot$/41, the second most metal-deficient BCG known after I 
Zw 18 (Figure 2b). As for the latter, we have used our own data instead of that 
of Skillman \& Kennicutt (1993).    
 The result is the same.  The scatter of the points is very small: 
$<$log N/O$>$ = --1.60$\pm$0.02 (Table 5). 
We do not believe that this very small scatter is the result of some unknown 
selection effect which would invariably pick out low-metallicity BCGs at the 
same stage of their evolutionary history. This is because of two reasons. First, 
we have plotted all the data we have, without any selection. Second, as 
discussed later, the scatter does increase substantially for higher-metallicity 
BCGs. There is no obvious reason why a selection effect would operate only on 
low-metallicity objects, but not on higher-metallicity ones.
  
As discussed by TIL95, the very small dispersion of the N/O ratio puts very 
severe constraints on time-delay models. A time delay between the primary 
production of oxygen by massive stars and that of nitrogen by intermediate-mass 
stars can be as large as 5$\times$10$^8$ yr, the lifetime of a 
2 -- 3 $M_\odot$ star.
This would introduce a significant ($\geq$ $\pm$ 0.2) scatter in N/O, as chemical 
evolution models by Pilyugin (1993) and Marconi et al. (1994)
show. We reiterate the conclusion of TIL95, that the small scatter of N/O in the 
most 
metal-deficient galaxies in our sample can be best understood if primary N in 
these galaxies is produced by massive stars ($M$ $>$ 9 $M_\odot$) only. 
As we shall see 
in section 5, intermediate-mass stars have not yet returned their 
nucleosynthesis products to the interstellar medium in these most metal-poor 
galaxies, because they have not had enough time to evolve.
 There is a further data point which 
we did not include in our sample, but which 
strengthens our results even more.  It comes from the western companion of SBS 
0335--052, the BCG SBS 0335--052 W, located in a common H I envelope with the 
former. This BCG was observed by Lipovetsky
et al. (1998) with the MMT and Keck telescopes to have oxygen
abundances in its two knots of 12 + log O/H = 7.22$\pm$0.03 and 7.13$\pm$0.07
respectively, with corresponding log N/O = --1.54$\pm$0.06 and --1.53$\pm$0.19,
consistent with our mean derived value of log N/O (Table 5). 

We have compared our results with those of other authors for BCGs in the 
range 12 + log O/H $\leq$ 7.6, using the compilation of Kobulnicky \& Skillman 
(1996) of the existing data for element abundances in BCGs. Two galaxies in 
their compilation show N/O ratios which are significantly lower than our mean 
value, by many times the dispersion of our sample. The first one is Tol 65, with 
12 + log O/H = 7.56 which has a very low log N/O = --1.79$\pm$0.20. This galaxy 
was observed nearly two decades ago (in 1980) by Kunth \& Sargent (1983). Its 
N/O ratio has large errors and should be redetermined more precisely. The second 
galaxy, CG 1116+51 with 12 + log O/H = 7.53 has log N/O = --1.68$\pm$0.11. 
However, Izotov \& Foltz (1998) have recently reobserved this 
galaxy with the MMT and found log N/O = --1.57$\pm$0.09, consistent with the 
mean value and small dispersion about the mean found here and in TIL95. 
Studies of southern BCGs by Campbell, Terlevich \& Melnick (1986) and Masegosa 
et al. (1994) also show a lower envelope for log N/O at around --1.6 
$\div$ --1.7. The 
very few galaxies with N/O below the envelope all have large observational
uncertainties.

   The situation changes appreciably for BCGs with 12 + log O/H $>$ 7.6. The 
scatter of the 
C/O and N/O ratios increases significantly at a given O abundance. This increase 
in the dispersion is best explained if, in addition to the production of carbon 
and primary nitrogen by massive stars during the starburst phase, there is an 
additional production of both these elements by intermediate-mass stars during 
the interburst quiescent phase. Within the framework 
of current nucleosynthesis theory (RV81, WW95), 
primary nitrogen is produced only by intermediate-mass stars, 
not by massive stars. In this case the productions of oxygen and nitrogen would 
be decoupled and in principle very low N/O values can be observed.
However, our observations do not confirm these expectations. Instead, Figure 2b 
shows a definite lower envelope for the N/O ratio, at the level set by 
primary nitrogen production in low-metallicity massive stars. 

    To summarize, we have arrived at the following important conclusions 
concerning the origin of nitrogen: (1) in very low-metallicity BCGs with 12 + 
log O/H $\leq$ 7.6, nitrogen is produced as a primary element by massive stars only.
Intermediate-mass stars have not had the time to evolve 
and release their nucleosynthesis 
products to the interstellar medium.
The massive stars set the level of log N/O to be 
at $\sim$ --1.60. This picture is the most reasonable one to account for the 
extremely small dispersion in log N/O ($\pm$ 0.02 dex) at a given O abundance.
(2) The value of log N/O increases above $\sim$ --1.60 along with the scatter at 
a given O abundance in BCGs with  7.6 $<$ 12 + log O/H $<$ 8.2. We interpret 
this increase in log N/O and its larger scatter as due to the additional 
contribution of primary nitrogen produced by intermediate-mass stars, on top of 
the primary nitrogen produced by massive stars. 

   Finally, we check whether the N/O ratios observed 
in more quiescent dwarf galaxies with less active star formation (van Zee et al. 
1997, 1998) are consistent with the scenario outlined above.
The H II regions in the latter have much lower excitation than 
those in BCGs, and the [O III] $\lambda$4363 emission line is not seen in many 
galaxies. In the framework of a time delayed nitrogen production model, we would 
expect lower values of N/O abundance ratios in quiescent dwarf galaxies than in 
BCGs as the bursts in the former are older (the H$\beta$ emission equivalent 
widths are smaller) and more oxygen has been released relative to nitrogen. Van 
Zee et al. (1997, 1998) do find some galaxies with very low log N/O $\leq$ --1.7. 
However, these low values are suspect and may be subject to systematic errors. 
There are several odd features concerning the galaxies with low N/O in the
sample of van Zee et al. (1997, 1998).
First, they were all observed in the same run, the mean value of 
their log N/O in 9 H II regions being --1.84, while that for the rest of the 
sample observed during 4 other runs is --1.52, in good agreement with the mean 
value found for BCGs. Second, the derived extinctions for the low N/O 
galaxies are either zero or systematically lower than that of other 
galaxies observed in different runs. 
The H II region UGC 5764-3 with the lowest log N/O = --2.02 has a 
H$\alpha$/H$\beta$ intensity ratio of only 2.3 , much lower than the theoretical 
recombination value of 2.7 -- 2.8. We conclude therefore that there is no strong 
evidence in the quiescent dwarf galaxy data  against a scenario of  primary 
nitrogen production by high-mass stars in the galaxy metallicity range  
12 + log O/H $\leq$ 7.6.

\subsection{Iron}

    The Fe/O abundance ratio also provides a very important constraint on the 
chemical evolution history of galaxies. TIL95 first discussed the iron abundance 
in BCGs. From their small sample of 7 low-metallicity BCGs, they found that 
oxygen in these galaxies is overproduced relative to iron, as compared to the 
Sun: [O/Fe] = 
log (Fe/O)$_\odot$ -- log (Fe/O) = 0.34$\pm$0.10. This value is in very good
agreement with  the [O/Fe] observed for Galactic halo stars (Barbuy 1988),
implying that the origin of iron in low-metallicity BCGs and in the
Galaxy prior the formation of halo stars is similar, and supporting the scenario 
of an early chemical enrichment of the galactic halo by massive stars.

    We have considerably increased the size of the sample of BCGs with iron 
abundance measurements. In 
Figure 2c we show [O/Fe] vs. 12 + log O/H for a total of 38 BCGs. It can be seen 
that, for all BCGs except one, [O/Fe] is above the solar value, reenforcing the 
conclusion of TIL95. The mean value of [O/Fe]
for the whole sample is 0.40$\pm$0.14 (Table 5). The only exception is the BCG 
SBS 0335--052 which has a negative [O/Fe], i.e. oxygen is underabundant with 
respect to iron as compared to the Sun. These odd abundances are not the result 
of observational errors as Izotov et al. (1997a) derived a similar result from 
independent MMT observations. The low value of [O/Fe] is more probably caused by 
the contamination and hence artificial enhancement of the nebular [Fe III] 
$\lambda$4658
emission line by the narrow stellar C IV $\lambda$4658 emission line produced in
hot stars with stellar winds. The presence of these stars in SBS 0335--052 is 
demonstrated by the detection of stellar Si IV $\lambda$$\lambda$1394, 1403 
lines with P Cygni profiles (Thuan \& Izotov 1997). A case in point which 
supports this contamination hypothesis is that of the NW component of I Zw 18. 
Izotov et al. (1997b) and Legrand et al. (1997) have discovered a Wolf-Rayet
population in this NW component with a significant contribution 
from WC4 stars to the C IV $\lambda$4658 emission. The spectrum by Izotov et al. 
(1997b) shows that a narrow emission line at $\lambda$4658 superposed on top of 
the broad C IV $\lambda$4658 bump produced by the WR stars. The use of
this narrow emission line to derive Fe abundance results in an artificially low 
[O/Fe] $\sim$ --0.4 for the NW component of I Zw 18. On the other hand, the SE 
component which does not possess Wolf-Rayet stars and hence has a nebular [Fe 
III] $\lambda$4658 emission line uncontaminated by narrow stellar C IV 
$\lambda$4658 emission, has a normal [O/Fe] $\sim$ 0.3, 
consistent with the value 
derived by ITL97 and with the mean value for the BCGs in our sample.
We have plotted in Figure 2c the [O/Fe] value for the SE component of I Zw 18. 
Another possible explanation for the abnormally low [O/Fe] in SBS 
0335--052 is the enhancement of the [Fe III] lines by supernova shocks.
Disregarding SBS 0335--052, Figure 2c shows that the
O/Fe ratio in BCGs is nearly constant, irrespective of the 
oxygen abundance, at a value $\sim$ 2.5 higher than in the solar neighborhood.

\subsection{Comparison of observational and theoretical nucleosynthetic
yields}

\subsubsection{Heavy element enrichment by massive stars}

     The remarkable constancy and small scatter of the C/O and N/O abundance 
ratios for the BCGs with oxygen abundance 12 + log O/H $\leq$ 7.6, and the 
similar behavior of the Ne/O, Si/O, S/O, Ar/O and [O/Fe] ratios with respect to 
the O abundance for all the
BCGs in our sample provide an unique opportunity to compare the observed
yields of massive stars with theoretical predictions, and put stringent 
constraints on  nucleosynthetic models of low-metallicity massive stars. This 
comparison has generally been made for stars in the Galaxy (Timmes et al. 1995; 
Samland 1998). However, the Galaxy is a complex evolved stellar system with the 
juxtaposition of many generations of stars. While its study gives the 
possibility to test models for a wide range of metallicities, the task is 
complex because the chemical enrichment is made not only by massive but also by 
intermediate and low-mass stars. Additionally,
heavy element abundance ratios in the Galaxy may be modified by dynamical
effects such as gas infall or outflow. BCGs are simple systems by comparison. As 
in the lowest-metallicity BCGs, all heavy elements are made by massive stars 
only, the chemical enrichment of BCGs is insensitive to infall or outflow of 
material, and is dependent only on the characteristics of the Initial Mass 
Function (IMF) and stellar yields. 

A first comparison of observed stellar yields by massive stars with theoretical 
calculations has been made by TIL95. Since our observational sample is much 
increased and new calculations (WW95) have appeared which 
cover a wider metallicity range (TIL95 had only solar metallicity models to 
compare with), we revisit the problem here. Table 6 shows
the heavy element-to-oxygen yield ratios as derived from the observed abundance 
ratios.
Since intermediate-mass stars contribute significantly to the synthesis of C and 
N in BCGs with 12 + log O/H $>$ 7.6, the observed
yield ratios $M$(C)/$M$(O) and $M$(N)/$M$(O) were derived using BCGs with
12 + log O/H $\leq$ 7.6, so that only the contribution from massive stars is 
taken into account.  For the other elements, the derived yields are simply the 
mean value of the abundance ratio for the whole sample. As for the theoretical 
yields given in Table 6, they are taken from WW95 and 
averaged over an IMF 
with a Salpeter slope in the stellar mass range 1 -- 100 $M_\odot$, for a heavy 
element mass fraction ranging from $Z$ = 0 to $Z_\odot$. The 2 horizontal solid 
lines in Figures 1 and 2 show, for all elements except N and Fe, 
the range of theoretical yield ratios predicted by WW95
as determined by their models with metallicities 0 and 0.01 $Z_\odot$.

 Inspection of Table 6 and Figures 1 and 2 shows that the heavy element yield 
ratios calculated by WW95 are generally not too far off from 
the yield ratios inferred from the observations.
The model which fits best the observed Ne/O and Si/O ratios (Figures 1a and 1b) 
has $Z$ = 0.  Models 
with $Z$ $>$ 0 predict a Ne abundance too low by a factor of $\sim$ 2, while the 
predicted Si abundance is too high by about the same factor. This 
anticorrelation can be explained by the fact that part of the Ne produced is 
consumed in the later stages of hydrostatic burning, synthesizing Si in 
particular. The observed S/O and Ar/O abundance ratios 
(Figures 1c and 1d) are best 
fitted by the $Z$ = 0.01 $Z_\odot$ and $Z$ = 0.1 $Z_\odot$ models which give 
nearly identical yields. 
The $Z$ = 0 model predicts yields too low by a factor of $\sim$ 1.3. As for C/O 
(Figure 2a), the $Z$ = 0 and $Z$ = 0.01 $Z_\odot$ models, where C is produced 
only by massive stars and not by intermediate-mass stars, bracket nicely the 
data for BCGs with 12 + log O/H $<$ 7.6.    
  It may seems surprising that the models which best fit elements such as Ne and 
Si have $Z$ = 0, considering that the BCGs in our sample all have $Z$ $\geq$ 0.02 
$Z_\odot$. However, these heavy element abundances characterize the ionized gas, 
not the stars which can have a much lower metallicity. 

Because the most abundant Ne isotope, $^{20}$Ne,
is synthesized during hydrostatic carbon burning, it is not sensitive to 
uncertainties in explosive nucleosynthesis models.    Si, S and Ar
yields in massive stars are, on the other hand, sensitive to the treatment of 
the explosion. Additionally, the production of these elements is sensitive to a 
variety of uncertain factors such as the rate of the  
$^{12}$C($\alpha$,$\gamma$)$^{16}$O process, the treatment of semiconvection,
the treatment of convection and convective overshoot mixing during the last
stages of shell oxygen burning, the density structure near the iron core, the
initial location of the mass cut, and the amount of mass that falls back in the
explosion (WW95). Given all these uncertainties, it is not so 
much the small discrepancies between theoretical and observational yield ratios 
that should be emphasized, but the overall good general agreement: it is    
remarkable that the abundance ratios inferred from the stellar yields by WW95 
do not differ from those observed in BCGs by large factors, but are 
invariably in the ballpark. 

    The calculated N and Fe yields constitute exceptions: they do not agree well 
with the data. TIL95 have already discussed the problem of iron, for which the
theoretical yields are $\sim$ 2 times greater than those inferred from the
observations. Theoretical calculations predict that iron is produced
during explosive nucleosynthesis by supernovae of both types I and II  
in nearly equal quantities. However, the progenitors of SNe II are short-lived
massive stars while the progenitors of SNe Ia are low-mass stars which explode 
only after 1 Gyr. Therefore, [O/Fe] is a good estimator of
the galaxy's age. The constancy of [O/Fe] and its high value 
in BCGs as compared to the Sun (Figure 2e) implies that iron in BCGs was 
produced only by
massive stars in type II supernovae. Because explosive nucleosynthesis models 
are sensitively dependent on the initial conditions of the explosion, the 
observed iron-to-oxygen abundance 
ratio can serve as a good discriminator between 
different models. TIL95 found that the models fit best the observations when the 
mass of the central collapsing core in the explosive synthesis is $\sim$ 10\% 
larger than the mass of the iron core. Since the mean value of [O/Fe] has not 
changed with the larger sample as compared to that found by TIL95, this 
conclusion still holds with the new data.
 
 The discrepancy between theory and observation is much more important for N. 
The N yield inferred from 
observations is 1 to more than 2 orders of magnitude larger than the theoretical 
yields of models with sub-solar metallicities (Table 6).
This is because conventional low-metallicity massive star models do not produce 
primary nitrogen. However, as noted by WW95, it is possible 
that in some massive stars, the convective
helium shell penetrates into the hydrogen layer with the consequent production
of large amounts of primary nitrogen. In fact, Timmes, Woosley \& Weaver (1995)
have found that the theoretical predictions for primary N production in massive 
stars with a large amount of convective overshoot are much more consistent with 
the observed [N/Fe] in low-metallicity halo stars as compared to conventional 
models, in which nitrogen is produced only as a secondary element, despite the  
unknown details of convective overshoot. 

\subsubsection{The role of intermediate-mass stars in heavy element
production}

   While a picture where all heavy elements in BCGs with 12 + log O/H $\leq$ 7.6
are produced in high-mass stars (HMS) only 
is consistent with the observations, the 
additional production of carbon and nitrogen by intermediate-mass stars (IMS) 
needs to be taken into account in BCGs with oxygen abundance 12 + log O/H $>$ 
7.6.

   As already discussed, it is commonly thought that primary nitrogen in 
low-metallicity BCGs is produced only by intermediate-mass stars (RV81, WW95). 
Additionally, some nitrogen is produced as a secondary element in both 
intermediate and high-mass stars. However, because production of 
secondary nitrogen drops as metallicity decreases, it is expected that the 
amount of secondary nitrogen is negligible compared to that of primary nitrogen 
in low-metallicity BCGs. As for carbon, it is believed to be produced as a 
primary element in all stars more massive than 1.5 $M_\odot$ (RV81, WW95). 

    We compare in Table 7 the observed C/O and N/O abundance ratios with  
theoretical predictions, taking into account the contributions of both high 
and  intermediate-mass stars. As before, all theoretical ratios are IMF-averaged 
values with the Salpeter slope of --2.35 and lower and upper mass limits of 1 
and 100 $M_\odot$ respectively.
 Theoretical yields for massive stars in the mass range
12 -- 40 $M_\odot$ and with a heavy element mass fraction $Z$ = 0.0002 are taken  
from WW95. Since production of primary nitrogen by massive 
stars is not considered by these authors, we adopt as the primary N yield by
massive stars that which is consistent with
the observed $<$log N/O$>$ = --1.60 for low-metallicity BCGs (Table 7 and 
solid horizontal line in Figure 2b). 
As for the C and N yields for intermediate-mass stars, they are taken from two 
different sets of models. The first set of models is from RV81. They are 
characterized by a stellar mass range from 3.5 to 7 $M_\odot$, 
a mass loss efficiency parameter on the asymptotic giant branch $\eta$ = 0.33, 
and a mixing length parameter $\alpha$ = 1.5. The other set of models 
is from HG97. They are characterized by $Z$ = 0.001 and 
a standard scaling parameter related to the efficiency of mass loss on the 
asymptotic giant branch $\eta$ = 1. We shall argue in Section 5 that the high 
value of [O/Fe] with respect to the Sun in the BCGs studied here implies that 
they are not older than $\sim$ 1 -- 2 Gyr.
Therefore, we have only considered yields from 
HG97 for stars with a lifetime less than 1 Gyr, i.e. with mass $\geq$ 2 
$M_\odot$.
We assume furthermore that oxygen is produced by massive stars only. We do not 
give in Table 7  the value of C/O in the case of C production by 
intermediate-mass 
stars only, as this situation is not realistic: C produced by longer-lived 
intermediate-mass stars is always accompanied by C produced by shorter-lived 
massive stars.

It is evident from Table 7 that there is general good agreement between 
observations and theory for C/O. 
We have already discussed the agreement for low-metallicity BCGs with oxygen 
abundance 12 + log O/H $\leq$ 7.6 where C is produced by massive stars only. The 
agreement is as good for higher-metallicity BCGs with 12 + log O/H $>$ 7.6, if  
C is produced in both massive and intermediate-mass stars. We have plotted in 
Figure 2a by a horizontal dashed line the theoretical C/O ratio calculated  
with models by WW95 for high mass stars and by RV81 for 
intermediate-mass stars. This value should be considered as an upper limit as 
the production of oxygen by massive stars in the current burst of star formation 
lowers the observed C/O ratios. The latter should lie between a lower limit set 
by primary C production by massive stars alone and whose possible range is shown 
by the two horizontal solid lines in Figure 2a, and that upper limit. The data 
points for BCGs with 12 + log O/H $>$ 7.6 do indeed scatter between these two 
limits as expected. Using the HG97 models would give a lower upper limit (by a 
factor of 1.6), but this is still consistent with the data given the 
observational uncertainties.

    The situation for nitrogen is more complex. 
If nitrogen is produced only by intermediate mass stars (and oxygen only by 
massive stars), then the RV81 and HG97 models predict respectively log N/O = 
--1.27 and --0.84. While these values are consistent with 
the largest N/O ratios observed for the BCGs in our sample with 12 + log O/H $>$ 
7.6 (Figure 2b), N production by intermediate-mass stars alone cannot explain, 
as discussed before, the very small dispersion of N/O ratios in BCGs with lower 
metallicities, so we do not consider this scenario further. 
We examine therefore the picture where nitrogen is produced by both high and 
intermediate-mass stars. If nitrogen is produced only as a secondary element
in massive stars, then the predicted log N/O is too small, $\leq$ --2.37 for 
massive star models with $Z$  $\leq$ 0.1 $Z_\odot$ (Table 6). We have thus to 
consider the situation where the nitrogen produced in 
massive stars is primary. In this
case, the combined (HMS + IMS) nitrogen and oxygen production gives log N/O = 
--1.11 and --0.77 for the RV81 and HG97 models respectively. As in the case of 
the C/O ratio, these values should be considered as upper limits. We have shown 
in Figure 2b by a dashed horizontal line the upper limit corresponding to the 
RV81 model. The data are completely consistent with the latter model: all the 
BCGs points fall within the dashed line and the lower limit set by primary N 
production in massive stars only (solid line). 
It is important to stress here that in this picture, no BCG can have a N/O (or 
C/O) ratio below the value set by massive star evolution, as indicated by the 
solid lines in Figure 2a (and 2b). 
We have already discussed in Sections 4.2 and 
4.3 that we know of no reliable data which contradict that statement.
  
In summary, the comparison of observational and theoretical yields shows a 
remarkably good general agreement in spite many uncertain parameters in the 
models. With the presently available data, the RV81 yields appear to give a 
slightly better fit to the data than the HG97 yields, although both sets are 
consistent with it within the observational uncertainties. It is also clear that 
further development of massive star nucleosynthesis theory is needed, especially
concerning nitrogen and iron productions. Because the theoretical yields 
of some elements are still so uncertain, we feel it is best to use, in computing 
chemical evolution models, the empirical yields 
derived from observations of low-metallicity BCGs, as summarized in Table 6.

\subsection{Evolution of the He abundance in BCGs}

   The analysis of the behavior of the heavy element abundance ratios as a 
function of oxygen abundance has shown that chemical enrichment proceeds 
differently in BCGs with low and high oxygen abundance. In BCGs with
12 + log O/H $\leq$ 7.6, high-mass stars are the main agents of chemical 
enrichment, the very small dispersion of the abundance ratios ruling out the 
time-delayed production of carbon and nitrogen by intermediate-mass
stars. On the other hand, in higher metallicity BCGs with 7.6 $<$ 12 + log O/H 
$<$ 8.2, the contribution of intermediate-mass stars to heavy element enrichment
is significant, as evidenced by an increase in the values and dispersions of the 
C/O and N/O ratios. One might expect therefore that the He enrichment history is 
also different in these two ranges of oxygen abundances, as helium is produced 
in different proportions by high and intermediate-mass stars.

   In Figure 3 we show the helium mass fraction $Y$ as a function of oxygen 
abundance for the galaxies in our sample. All of them possess accurate He 
abundance determination as the present sample is precisely the one used by 
Izotov \& Thuan (1998b) to derive the primordial helium abundance $Y_p$. 
It is usual practice to extrapolate the $Y$ versus O/H and $Y$ versus N/H linear 
regressions to O/H = N/H = 0 to derive $Y_p$ (Peimbert \& Torres-Peimbert 1974,
1976; Pagel, Terlevich \& Melnick 1986). The $Y$ versus O/H regression line is 
decribed by  
\begin{equation}
Y = Y_p + \frac{dY}{d({\rm O/H})} ({\rm O/H}).      \label{eq:YvsO}
\end{equation}

The dotted line in Figure 3 represents the best fit regression line as derived 
by Izotov \& Thuan (1998b), with d$Y$/d$Z$ = 2.4  which corresponds to
d$Y$/d(O/H) = 45. 

We now compare this best fit slope with the ones predicted in various models.
In the scenario of element production in high-mass stars only for BCGs
with 12 + log O/H $\leq$ 7.6 or O/H $\leq$ 4$\times$10$^{-5}$, the predicted 
slope (WW95) is much shallower, d$Y$/d$Z$ = 0.94 (Table 7). 
The solid line in Figure 3 
has this slope (we adopt $Y_p$= 0.244, Izotov \& Thuan 1998b). It can be seen 
that it fits quite well the lowest-metallicity points, and is in fact nearly 
indistinguishable from the best fit line. At higher oxygen abundances, the slope 
steepens because of the additional production of helium in intermediate mass 
stars and takes the values d$Y$/d$Z$ = 1.66 and 1.54 for the RV81 and HG97 
yields respectively (Table 7). 
These values are lower than the best fit slope d$Y$/d$Z$ = 2.4$\pm$1.0, but are 
consistent with it within the errors. The dashed line in Figure 3 has the slope 
derived with the RV81 yields. Taking into account the error bars, it fits the 
observations quite well. We shall need to supplement our BCG sample with more 
high-metallicity BCGs to determine d$Y$/d$Z$ better, and reduce the 
observational uncertainties in the individual He determinations, before we can 
ascertain whether there is any difference between the best fit slope and the one 
derived from theory. Given the present data, we conclude
that our proposed scenario -- He production in high-mass stars only for  
galaxies with 12 + log O/H $\leq$ 7.6, and He production by both high and 
intermediate-mass stars for higher metallicity galaxies -- 
is in good agreement with the observations. If we assume that 25\% of the oxygen 
produced is lost by the galaxy due to supernova-driven winds, then d$Y$/d$Z$ 
is nearly unchanged at low metallicities because both oxygen and helium are 
produced in massive stars. However, when intermediate-mass stars play 
a role in producing He (but not O), the slope steepens, becoming 1.89 instead of 
1.66. 

The above analysis implies that a slope change may be expected for the $Y$ 
versus O/H linear regression at 12 + log O/H $\sim$ 7.6, and that fitting both 
the low and high metallicity ranges by the same slope may introduce some 
systematic underestimation of $Y_p$. If this is the case, it is perhaps best to 
derive $Y_p$ not by a regression fit, but by taking the mean $Y$ 
of the most metal-deficient galaxies. ITL97 and Izotov \& Thuan (1998b) did 
indeed find that the mean $Y$ of the two most metal-deficient BCGs known, I Zw 
18 and SBS 0335--052, is higher by 0.001 than the value derived from a linear 
regression fit to the whole sample ($<$$Y$$>$ = 0.245$\pm$0.004 
instead of $Y_p$ = 0.244$\pm$0.002). 
However, as shown in Figure 3, the difference between the regression line 
derived by fitting the data with the same slope over the whole metallicity range
and that expected for He production by massive stars only, is 
small and far below the observational uncertainties. 
We conclude therefore that, given the present quality of the data, the method of
using a single linear regression (Eq. \ref{eq:YvsO}) with the same d$Y$/d$Z$ 
slope over the whole metallicity range is perfectly adequate for the 
determination of the primordial helium abundance.

We next examine the behavior of He with respect to C in the context of our 
favored model: He and heavy-element production by high-mass stars only in BCGs 
with 12 + log O/H $<$ 7.6, and by both high-mass and intermediate-mass stars in 
higher metallicity BCGs. 
In Figure 4 we show the dependence of the helium mass fraction $Y$ on the 
carbon abundance C/H for those BCGs in our sample for which both of these 
quantities are known. Admittedly, the total number of such galaxies is very 
small (only 4), too small to draw any definite conclusion on the slope of the 
$Y$ -- C/H relation. The solid line shows the expected relation when He and C 
are only produced by high-mass stars, and the dashed line that expected when 
both high and intermediate-mass stars contribute. The theoretical IMF-averaged 
yields are taken from WW95 for high-mass stars and from RV81 
for intermediate-mass stars. Despite
the very small number of data points, it can be seen that the theoretical 
predictions agree well with the observational data. We give in Table 7 the 
theoretical slopes d$Y$/d(C/H) in both situations (HMS and IMS+HMS), for two 
sets of yields for intermediate-mass stars taken from RV81 and HG97.
While the slopes in the case of HG97 yields are nearly the same in both 
situations, the slope with RV81 yields is shallower when both high and 
intermediate-mass stars produce He and C than when only high-mass stars are 
responsible for the nucleosynthesis.

\section{CHEMICAL CONSTRAINTS ON THE AGE OF BCGs}

   Because O, Ne, Si, S and Ar are made in the same high-mass stars, their 
abundance ratios with respect to O are constant and not sensitive to
the age of the galaxy. By contrast, C, N and Fe can be produced by both high 
and lower-mass stars, and their abundance ratios with respect to O can 
give important information on the evolutionary status of BCGs. 

   The constancy of [O/Fe] for the BCGs in our sample and its high
value compared to the Sun (Figure 2c) suggests that all iron was produced by 
massive stars, i.e. in
SNe II only. Since the time delay between iron production from SNe II and SNe Ia is about 1 -- 2 Gyr, it is likely that BCGs with oxygen
abundance less than 12 + log O/H $\sim$ 8.2 are younger than
1 -- 2 Gyr, assuming that abundances measured in the supergiant H II regions are 
representative for the whole galaxy. We cannot put more stringent constraints 
on the star formation history of BCGs with the [O/Fe] abundance ratio because  
intermediate-mass stars do not produce oxygen and iron.

   We next use the behavior of the C/O and N/O ratios as a function of oxygen 
abundance to constrain the age of BCGs. As discussed previously, this behavior 
is very different whether the BCG has 12 + log O/H smaller or greater than 7.6.
The C/O and N/O abundance ratios in BCGs with 12 + log O/H $\leq$ 7.6 are 
independent of the oxygen abundance and show a very small scatter about the mean 
value. We have argued that this small scatter rules out any time-delay model 
in which O is produced first by massive stars and C and N are produced later by 
intermediate-mass stars, and supports a common origin of C,
N and O in the same first-generation massive stars. Thus, it is very 
likely that the presently observed episode of star formation in BCGs with 12 + 
log O/H $\leq$ 7.6 is the first one in the history of the galaxy and the age of 
the oldest stars in it do not exceed $\sim$ 40 Myr, the lifetime of a 
9 $M_\odot$ star. 

    The conclusion that BCGs with $Z$ $\leq$ $Z_\odot$/20 are young is 
supported by the analysis of {\sl HST} WFPC2 images of some of these galaxies.
Hunter \& Thronson (1995) have concluded that the blue colors of the underlying 
diffuse extended emission in I Zw 18 ($Z_\odot$/50) are consistent with those 
from a sea of unresolved B and early A stars, with no evidence for stars older 
than $\sim$ 10$^7$ yr.  Izotov et al. (1997a) and 
Thuan, Izotov \& Lipovetsky (1997) have also found in SBS 
0335--052 ($Z_\odot$/41), after removal of the gaseous emission, an extremely 
blue extended underlying stellar component with age less than 100 Myr. In 
the same manner, SBS 1415+437 ($Z_\odot$/21, Thuan, Izotov \& Foltz 1998), 
T1214--277 and Tol 65 (respectively $Z_\odot$/21 and $Z_\odot$/22, Izotov, 
Thuan \& Papaderos 1998) show, after subtraction of the gaseous emission, a very blue extended emission consistent with an underlying stellar population not 
older than 100 Myr.

   The situation changes for BCGs with $Z$ $>$ $Z_\odot$/20. The scatter of the 
C/O and N/O ratios increases significantly at a given O abundance, which we 
interpret as due to the additional production of primary N by intermediate-mass 
stars, on top of the primary N production by high-mass stars. 
Thus, since it takes 
at least  500 Myr (the lifetime of a 2 -- 3 $M_\odot$ star) for C and N to be 
produced by 
intermediate-mass stars, BCGs with 12 + log O/H $>$ 7.6 must have had several 
episodes of star formation before the present one and they must be at least 
older than $\sim$ 100 Myr. This conclusion is in agreement with photometric 
studies of these higher metallicity BCGs which, unlike their very 
low-metallicity counterparts, have a red old instead of a blue young underlying 
stellar component (Loose \&
Thuan 1985; Papaderos et al. 1996; Telles \& Terlevich 1997).   

 In summary, the study of heavy element abundances in BCGs leads us to the
following timeline for galaxy evolution: a) all objects with 12 + log O/H $\leq$
7.6 began to form stars less than 40 Myr ago; b) after 40 Myr, all galaxies have
evolved so that 12 + log O/H $>$ 7.6; c) by the time intermediate-mass stars have
evolved and released their nucleosynthetic products (100--500 Myr),
all galaxies have become enriched to 7.6 $<$ 12 + log O/H $<$ 8.2.  The delayed
release of primary N at these metallicities greatly increase the
scatter in the N/O abundance ratio; d) later, when galaxies get enriched to 12
+ log O/H $>$ 8.2, secondary N production becomes important.

\section{COMPARISON WITH DAMPED LYMAN-$\alpha$ SYSTEMS}

    Damped Ly$\alpha$ systems are believed to be young disk galaxies in their 
early stages of evolution. They are extremely metal-deficient, their heavy 
element abundances ranging between $Z_\odot$/300 and $Z_\odot$/10 (Pettini
et al. 1994; Wolfe et al. 1994; Lu et al. 1996). The large light-gathering
power of the 10 m Keck telescope has made it possible to study 
their elemental abundance ratios (Lu et al. 1996), thus revealing many 
similarities between these ratios and those found in BCGs.
 A direct comparison, however, is not always possible. The Ne and Ar emission 
lines are present in the spectra of BCGs, but the absorption lines of the same 
noble elements are absent in the spectra of Ly$\alpha$ galaxies. On the other 
hand, the abundances of several elements in the iron-peak group have been 
measured in damped Ly$\alpha$ galaxies, but only iron abundance measurements are 
available for BCGs.
There exists only lower limits for the carbon and oxygen abundances in damped 
Ly$\alpha$ systems as the O I $\lambda$1302 and
C II $\lambda$1334 absorption lines are strongly saturated. 
In the case of the elements for which a direct comparison can be made, Lu et al. 
(1996) have shown that some $\alpha$-elements such as Si and S are overabundant 
with respect to iron as compared to the Sun. This 
is true as well for BCGs (and for halo stars) since in these objects, Si and S
are normal with respect to O while O is overabundant with respect to Fe, by
comparison with the Sun. 
On the other hand, the iron-peak 
element abundance ratios are nearly solar as expected.

   The N/O ratios seem at first glance to constitute the major difference. The 
N/O ratios in damped Ly$\alpha$ systems appear to be significantly lower than 
those measured in our low-metallicity BCGs.  Pettini, Lipman \& Hunstead (1995) 
have found in one damped Ly$\alpha$ galaxy an upper limit log N/O $\leq$ --2.12 
, --0.52 dex lower than the mean log N/O =
--1.60 for the most metal deficient BCGs in our sample, a value which, we have 
argued, is set by primary N production in massive stars and constitutes a lower 
limit to log N/O in BCGs. Lu, Sargent \&
Barlow (1998) have measured N/Si abundance ratios in 15 damped Ly$\alpha$
galaxies, and have set upper limits for log N/O ranging from from --1.2 to 
--2.1. Until now, nearly all log N/O measurements lower than the BCG lower limit 
of --1.60 in damped Ly$\alpha$ systems are 
upper limits. There is however one exception, the 1946+7658 system where Lu et 
al. (1998) did measure an actual value, [N/Si] = --1.70 which translates to log 
N/O = --2.6, or 1.0 dex lower than the BCG lower limit. From the N/O and N/Si 
measurements, Pettini et al. (1995) concluded that both
primary and secondary nitrogen production are important over the whole range of  
metallicity measured in damped Ly$\alpha$ systems,
while Lu et al. (1998) favor the time-delayed primary nitrogen production
by intermediate-mass stars.

   We find such a large difference between the N/O ratios measured in BCGs and 
those in damped Ly$\alpha$ galaxies to be very puzzling. If we believe the 
physics of star formation, the stellar IMF and the nucleosynthesis processes in 
stars at a given metallicity, to be universal and independent of cosmic epoch, 
then there must be some systematic differences in the way in which 
the N/O ratios are derived in damped Ly$\alpha$ systems as compared to 
those in BCGs.
The derivation of N/O ratios in BCGs is straightforward enough.
The photoionized H II region models are simple and well defined, the nebular 
emission lines are strong and their intensities can be measured with precision. 
There is no hidden assumption in the path from line intensities to N and O 
abundances.
On the other hand, there is a basic assumption in the derivation of N/O in 
damped Ly$\alpha$ systems which we would like to discuss. It is generally 
assumed that absorption lines in damped Ly$\alpha$ systems originate in the H I 
clouds in the disk of the galaxy. In this case, 
only low ionization species are expected and correction factors for unseen 
stages of ionization are close to unity (Viegas 1995).
We suggest here that this basic assumption may not be valid. We may  
expect that in high-redshift gas-rich disks there is ongoing star formation 
giving birth to massive stars which ionize the H I gas and create H II regions
and diffuse ionized gas. 
There is thus a non-zero probability for the line of sight to cross  diffuse ionized gas
 which is ubiquitous and has a large covering 
factor in gas-rich galaxies and/or HII regions, 
so that absorption lines would originate not in neutral but 
ionized gas. Supporting this hypothesis is the fact that many of the spectra of 
Lu et al. (1996) do show absorption lines of high ionization species such as Al 
III, C IV and Si IV, which are usually present in ionized gas regions. 

Lu et al. (1996) have dismissed the ionized gas hypothesis by the following two 
arguments. First, the absorption profile of the high-ionization species Al III 
is similar to those of lower ionization species, and hence both types of
species must be produced in the same physical region. The latter must be
mostly neutral because of the large observed H I column densities. In that case,
Al III comes from an ionized shell surrounding the H I gas.
Second, the Al II lines are always much stronger than the 
Al III lines, so that $N$(Al II) $\gg$ $N$(Al III) implying that most of the gas 
where the absorption arises is neutral. 
The last argument is not airtight.
Supposing that all absorption lines originate not in H I but H II gas, we 
have constructed with the radiative-collisional equilibrium code 
CLOUDY (Ferland 1993) a series of spherically symmetric 
H II region models with the ionization parameter log U in the range --0.4 to 0
which is typical for HII regions, and ionizing stars with 
 effective temperatures between 40 000 and 50 000 K and a metallicity of one-tenth
 solar. For these models, typical radially-averaged column densities are 
 $N$(H II) = 10$^{19}$--10$^{21}$ cm$^{-2}$ and 
 $N$(Al II) = 10$^{12}$--10$^{14}$ cm$^{-2}$.
 The models give invariably  
$N$(Al II) $>$ $N$(Al III) even when the hydrogen gas is totally ionized. 
The profile similarities 
noted by Lu et al. (1996) between Al III and Al II can indeed be explained  
by the formation of absorption lines in the same physical region, but the 
latter can be H II rather than H I gas. 
If the gas is ionized, the correction factors 
for unseen stages of ionization are dependent on the parameters of the 
particular H II region model and will be higher than for those in a H I cloud. 
The determination of abundances is more uncertain because of the lack of 
information on the column density of ionized hydrogen. 
However, the abundance ratios for some elements will not change greatly when the 
absorption lines are assumed to originate in the H II instead of the H I gas. 
This is because the abundances of these elements are derived from
column densities of singly ionized ions, e.g. C$^+$, Si$^+$, S$^+$,
Fe$^+$. Photoionized H II region models (Stasi\'nska 1990) predict for these 
ions similar correction factors,
so that their ratios are close to unity and the element abundance ratios are 
roughly equal to the singly ionized ion
abundance ratios. However, the situation changes dramatically 
when we compare abundances of elements derived from column densities
of ions in different stages of ionization. In those cases, the ratio of the 
ionization correction factors for different elements can be very far from unity, 
and the abundance ratios very different from the 
ion abundance ratios.  Such is the 
case for the N/Si abundance ratio. While the silicon column density is derived 
from 
the Si$^+$ absorption line equivalent widths, the nitrogen column density is
derived from the N$^0$ absorption lines. In the H I cloud model 
\begin{equation}
\frac{\rm N}{\rm Si} = \frac{\rm N^0}{\rm Si^+}.  \label{eq:HI}
\end{equation}
The situation is however totally different if the H II region model is adopted.  
In Figure 5 we show the correction factors ICF(N/Si) as a function of the 
fraction of O$^+$ ions $x$(O$^+$) = O$^+$/O, for the set of H II region models 
of Stasi\'nska (1990).
The correction factors ICF(N/Si) are weakly dependent on $x$(O$^+$) and have a 
lower envelope at a value $\sim$ 10. Hence, for the H II region model
\begin{equation}
\frac{\rm N}{\rm Si} = {\rm ICF}\left(\frac{\rm N}{\rm Si}\right) 
\frac{\rm N^0}{\rm Si^+} \geq 10\frac{\rm N^0}{\rm Si^+}.  \label{eq:HII}
\end{equation}
This factor of $\sim$ 10 is just about the one needed to bring the N/O ratio 
measured in the 1946+7658 damped Ly$\alpha$ system to the mean value of N/O 
obtained for low-metallicity BCGs.

Instead of the N/Si ratio, Pettini et al. (1994) have measured directly the N/O 
ratio in the damped Ly$\alpha$ systems. This method is in principle subject to 
less uncertainties because the abundances of both elements in the H II region 
are 
derived from column densities of neutral species. We have run the CLOUDY code 
for several spherically symmetric H II region models, varying the ionization 
parameter and the temperature of the ionizing radiation. These calculations show 
that the fractions of neutral nitrogen and neutral oxygen in the H II
region along the line of sight are nearly the same, i.e.
\begin{equation}
\frac{\rm N}{\rm O} \approx \frac{\rm N^0}{\rm O^0}.  \label{eq:NO}
\end{equation}
However, the O abundance is not known with good precision. The O I $\lambda$1302 
absorption line is saturated and the O abundances derived by Pettini et al. 
(1995) by fitting the saturated O I profiles have so large errors that they span 
 3 orders of magnitude. Furthermore, the very method of using saturated 
absorption lines to derive column densities and abundances is questionable 
(Pettini et al. 1995; Pettini \& Lipman 1995; Lu et al. 1998).

  We stress therefore that there is a great uncertainty in the derived N/O 
abundance ratios in damped Ly$\alpha$ absorption systems. This uncertainty comes 
in large part from our ignorance of the nature of the absorbing medium, whether 
it is neutral or ionized gas. If the absorbing system is an H II region instead 
of a H I cloud, the derived N/O abundance ratio may increase by a factor of 10 
or more if the Si$^{+}$ lines are used instead of neutral oxygen lines, 
which would bring the N/O ratios derived in damped Ly$\alpha$ systems 
in closer agreement with those found in low-metallicity BCGs. Another large 
source of uncertainty comes from the derivation of abundances from saturated O I 
$\lambda$1302 absorption lines. If in the future, there is evidence that the 
absorption lines do come from a neutral medium, and better O determinations in   
damped Ly$\alpha$ systems still give a N/O ratio a whole order of magnitude 
lower than the value found in low-metallicity BCGs, then we would have to  
conclude that star formation and metal dispersal histories in damped 
Ly$\alpha$ galaxies are very different from those in BCGs.

\section {SUMMARY AND CONCLUSIONS}

    The present study is a continuation and extension of the one by TIL95 on 
heavy element abundances in very metal-deficient environments, with considerably 
more data. We derive here the abundances of N, O, Ne, S, Ar and Fe in a sample 
of 54 supergiant H II regions in 50 low-metallicity blue compact galaxies with 
oxygen abundance in the range 7.1 $<$ 12 + log O/H $<$ 8.3. The objects in this 
sample all possess very high signal-to-noise ratio spectra which have been 
obtained for the determination of the primordial helium abundance. This allows 
us to measure line intensities with a high precision, and properly correct them 
for interstellar extinction and underlying stellar hydrogen Balmer absorption. 
In addition, we redetermine the carbon and silicon abundances in some 
BCGs with the use of {\sl HST} UV and 
optical archival spectra, supplemented in a few cases by ground-based optical 
spectroscopic observations to derive accurate electron temperatures.

We have obtained the following results:

1. As in TIL95, the $\alpha$-elements-to-oxygen Ne/O, Si/O, S/O and 
Ar/O abundance ratios show no dependence on oxygen abundance over the whole 
range of metallicities studied here. Furthermore, these ratios are about the 
same as those found in halo stars and high-redshift damped Ly$\alpha$ galaxies, 
and they have approximately the solar values. This result is to be expected from 
stellar nucleosynthesis theory as oxygen and all $\alpha$-elements are produced 
by the same massive stars.

2. We rederive the carbon-to-oxygen abundance ratio in both NW and SE components 
of I Zw 18, the most metal-deficient galaxy known. Our values of log C/O = 
--0.77$\pm$0.10 for the NW component and
--0.74$\pm$0.09 for the SE component are in excellent agreement with those  
predicted by the theory of massive star nucleosynthesis, but are
significantly lower (by $\sim$ 0.2 dex) than those derived by Garnett et al.
(1997). The main source of the differences comes from the adopted electron 
temperatures. We use higher electron temperatures (by 1900 K and 2300 K 
respectively for the NW and SE components), as derived from recent 
MMT spectral observations (Izotov et al. 1997b) in apertures which match more 
closely those of the {\sl HST} FOS observations used to obtain C abundances.
With these lower C/O ratios, I Zw 18 does not stand apart anymore from the other 
low-metallicity BCGs. An earlier lower-mass carbon-producing stellar population 
need not be invoked, and I Zw 18 can still be a 
``primeval" galaxy undergoing its 
first burst of star formation. 

3. The C/O abundance ratio is constant, independent of the O abundance in BCGs 
with 12 + log O/H $\leq$ 7.6 ($Z$ $\leq$ $Z_\odot$/20), with a mean value log 
C/O 
= --0.78$\pm$0.03 and a very small dispersion about the mean. This result is based 
on a small sample of 4 data points in 3 galaxies and needs to be strengthened 
by more C/O measurements in extremely metal-deficient galaxies.
The constancy of the C/O ratio suggests that carbon in these galaxies is produced in 
the same stars that make O, i.e. only in massive 
stars ($M$ $>$ 9 $M_\odot$). By 
contrast, the C/O ratio in BCGs with higher oxygen abundance (12 + log O/H $>$ 
7.6) is significantly higher, with a mean value log C/O = --0.52$\pm$0.08
and a larger dispersion about the mean. This enhanced C/O ratio and the larger
scatter at a given O/H are likely the result of additional carbon production by 
intermediate-mass (3 $M_\odot$ $\leq$ $M$ $\leq$ 9 $M_\odot$) stars, 
on top of the carbon production by high-mass stars.

4. TIL95 showed that the N/O abundance ratio has a very small scatter in BCGs 
with 12 + log O/H $\leq$ 7.6, with a mean value log N/O = --1.58$\pm$0.02. We 
have added here a very low-metallicity object, SBS 0335--052 with $Z_\odot$/41, 
and the results do not change. As discussed by TIL95, such a small dispersion of 
the N/O ratio is not consistent a time-delayed production of primary nitrogen by 
longer-lived intermediate-mass stars in these 
extremely metal-deficient galaxies. The small dispersion 
at very low metallicities can only be explained 
 by primary nitrogen production in short-lived massive stars. As in the case 
of C, galaxies with 12 + log O/H $>$ 7.6 show a considerably larger scatter of 
N/O ratios, with a mean value log N/O = --1.46$\pm$0.14. Again, the larger 
scatter can be explained by the addition of primary nitrogen production in 
intermediate-mass stars, on top of that in massive stars. None of our galaxies 
has log N/O below --1.67, implying that the lower envelope of the N/O 
distribution is set by the production of primary nitrogen in massive stars over 
the whole range of metallicities studied here, 7.1 $<$ 12 + log O/H $<$ 8.2.  
While a detailed scenario of primary nitrogen production by massive stars is yet 
to be developed, the mean log N/O obtained here for the most metal-deficient 
BCGs will put strong constraints on any future theory.

5. The Fe/O abundance ratio in our BCGs is $\sim$ 2.5 times lower than in the 
Sun, with a mean [O/Fe] = 0.40$\pm$0.14.
Again, it does not show any dependence on oxygen abundance, in agreement
with TIL95. This implies that iron in BCGs was synthesised during the explosive 
nucleosynthesis of SNe II. Only one BCG bucks the trend: SBS 0335--052 has a 
Fe/O abundance ratio higher than the solar ratio. We argue
that the high abundance of Fe in this BCG may not be real, and may be caused by 
the contamination of the nebular [Fe III] $\lambda$4658 emission line by the 
narrow stellar C IV $\lambda$4658 produced in expanding envelopes of hot massive 
stars. 

6. Comparison of theoretical heavy element yields for low-metallicity
massive stars (WW95) with those inferred from  
observations of BCGs shows that the theory of massive star nucleosynthesis
is generally in good shape. The major problems are with iron and
nitrogen yields. The predicted iron-to-oxygen yield ratio 
is a factor of $\sim$ 2 larger than the observed ratio. Models of primary 
nitrogen production by intermediate-mass stars cannot reproduce the N/O ratios 
observed in very low-metallicity BCGs, and conventional models of 
low-metallicity massive stars do not allow for primary nitrogen production. 
Until further developments of the theory of massive star nucleosynthesis resolve 
these disagreements, it is best to use the observed yields given in Table 6 for  
studies of the early chemical evolution of galaxies.

7. Helium follows the same pattern as carbon and nitrogen. The helium abundance 
is constant within the observational uncertainties for BCGs with 
12 + log O/H $\leq$ 7.6. Again we interpret this constancy by He being produced 
in high-mass stars only. At higher oxygen abundances, the mean and dispersion 
increase, and we again interpret this increase as the addition of helium 
production by intermediate-mass stars to that by massive stars. Although the 
sources of helium production appear to be different in low and high-metallicity 
BCGs, implying slightly different dependences of the helium mass fraction $Y$ on 
oxygen abundance above and below 12 + log O/H = 7.6, we find that the commonly 
used $Y$ versus O/H linear regression over a large range of oxygen abundances
to determine the primordial helium abundance, is a good approximation.

8. The constancy of the C/O, N/O and [O/Fe] ratios in BCGs with
12 + log O/H $\leq$ 7.6 argues strongly for the production in the most
metal-deficient galaxies, of these elements
by massive stars only. Intermediate-mass stars have not yet returned 
their nucleosynthesis products to the interstellar medium 
in these most metal-poor galaxies because they have not 
had enough time to evolve. 
This allows us to date these 
galaxies. Galaxies with 12 + log O/H $\leq$ 7.6 are young, in the sense that 
they have experienced their first episode of star formation not more than 
$\sim$ 40 Myr ago, the lifetime of a 9 $M_\odot$ star. 
This derived young age is 
in agreement with the results obtained by recent photometric studies with {\sl 
HST} images of some of the most metal-deficient galaxies discussed here.

9. The study of heavy element abundances in BCGs leads us to the
following timeline for galaxy evolution: a) all objects with 12 + log O/H $\leq$
7.6 began to form stars less than 40 Myr ago; b) after 40 Myr, all galaxies have
evolved so that 12 + log O/H $>$ 7.6; c) by the time intermediate-mass stars have
evolved and released their nucleosynthetic products (100--500 Myr),
all galaxies have become enriched to 7.6 $<$ 12 + log O/H $<$ 8.2.  The delayed
release of primary N at these metallicities greatly increase the
scatter in the N/O abundance ratio; d) later, when galaxies get enriched to 12
+ log O/H $>$ 8.2, secondary N production becomes important.

10.  The N/O abundance ratios derived for BCGs are apparently larger by a factor 
of up to $\sim$ 10 than those obtained for high-redshift damped Ly$\alpha$
galaxies (Lu et al. 1996). We suggest here that this discrepancy may not be 
real. Contrary to the situation for BCGs where the N/O ratios are very well 
determined, the N/O ratios derived for damped Ly$\alpha$ systems are highly 
uncertain because of the unknown physical conditions in the interstellar medium
of high-redshift systems. It is generally assumed that the absorption lines in 
these systems originate in the H I gas. In that case, the ionization correction 
factors for the low-ionization species used for abundance determination are 
close to unity. However, if we assume that the absorption lines originate not in 
neutral but ionized gas, the correction factors can be $\geq$ 10, 
and the 
derived N/O abundance ratios in the damped Ly$\alpha$ systems can be as high as 
those derived in BCGs. 

\acknowledgements
Y.I.I. thanks the warm hospitality of the Astronomy Department of the University 
of Virginia. This international collaboration was made possible thanks to the  
partial financial support of NSF grant AST-9616863. 


\clearpage

\figcaption[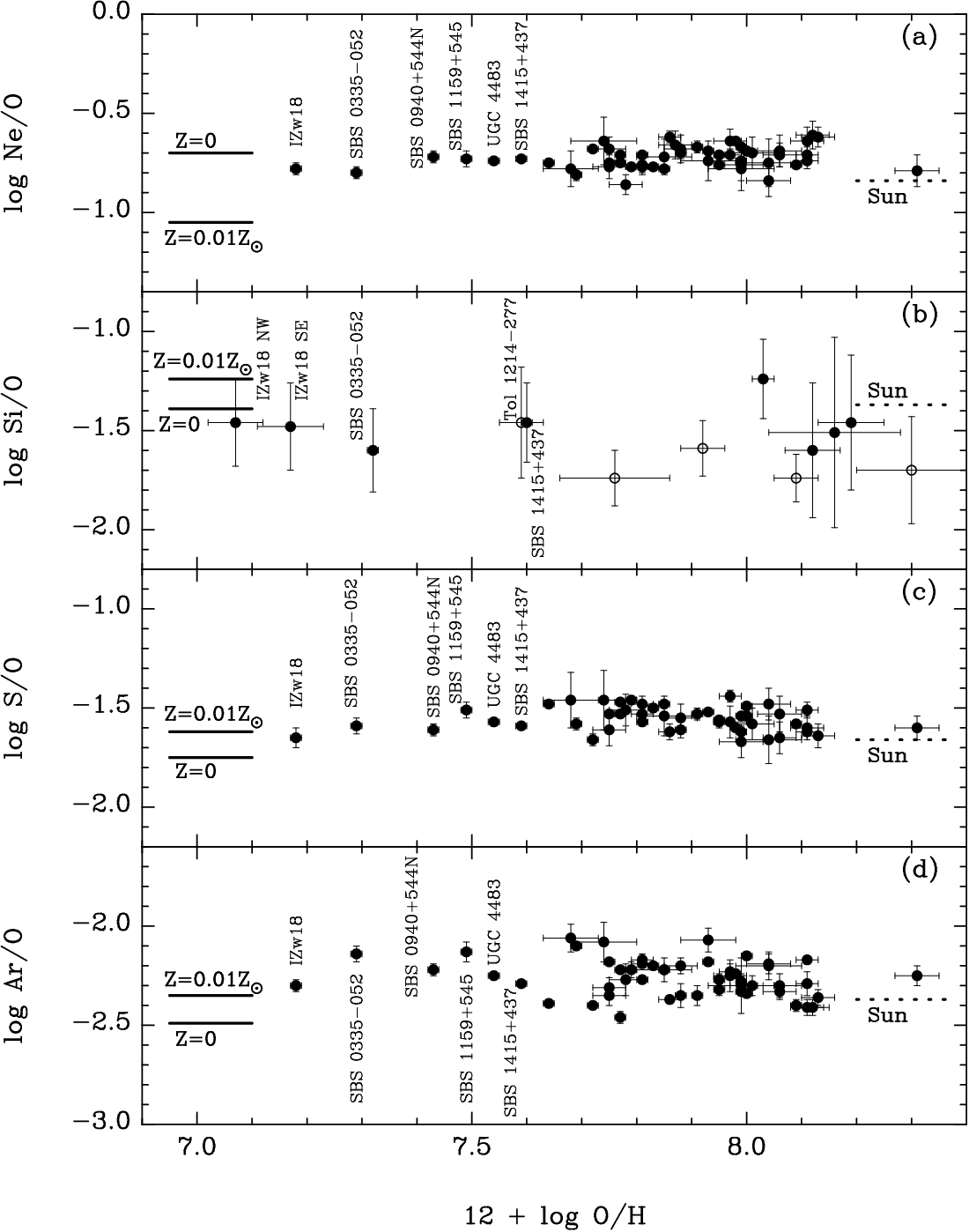]{ $\alpha$-element-to-oxygen Ne/O, Si/O, S/O
and Ar/O abundance ratios as a function of oxygen abundance. Except for Si, the 
galaxy sample shown here is the same as the one used by Izotov \& Thuan (1998b) 
to derive the primordial helium abundance. Concerning Si (Figure 1b), we have 
shown by filled
circles the Si/O abundance ratios reanalyzed here using the {\sl HST} UV and 
optical data by Garnett et al. (1995b), Kobulnicky et al. (1997), Kobulnicky \& 
Skillman (1998) and our optical data obtained with the Multiple Mirror 
Telescope. 
Other data from Garnett et al. (1995b), 
Kobulnicky et al. (1997), Kobulnicky \& Skillman (1998) are shown by open 
circles. All points with oxygen abundance less than 1/20 that of the Sun are 
labeled. The solid horizontal lines show the theoretical ratios given by massive 
star nucleosynthesis models by WW95 with metallicities 0 and 
0.01 $Z_\odot$ (Table 6). The dashed horizontal line shows the solar ratio.}

\figcaption[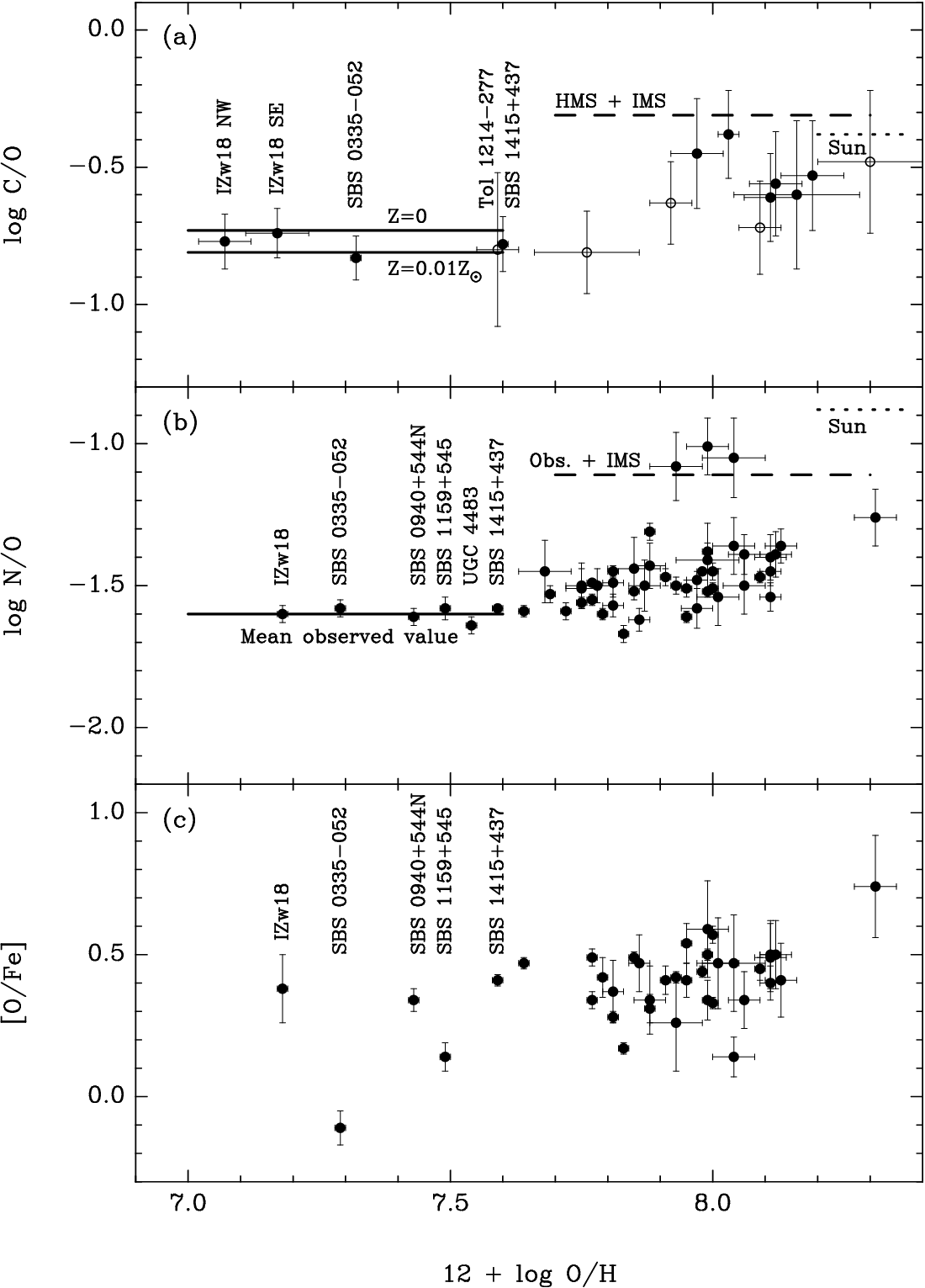]{The C/O, N/O and [O/Fe] abundance ratios as a function of 
oxygen abundance. Except for C, the galaxy sample shown here is the same as the 
one used by Izotov \& Thuan (1998b) to derive the primordial helium abundance. 
Concerning C (Figure 2a) we show by filled
circles the C/O abundance ratios rederived here using the {\sl HST} UV and 
optical data by Garnett et al. (1995a), 
Kobulnicky et al. (1997), Kobulnicky \& Skillman (1998) and our optical  
data obtained with the Multiple Mirror Telescope. 
Other data from Garnett et al. (1995a), 
Kobulnicky et al. (1997), Kobulnicky \& Skillman (1998) are shown by open 
circles. Note the very small dispersion of the C/O and N/O ratios for BCGs with
12 + log $\leq$ 7.6 (the labeled points). For C/O (Figure 2a), 
the 2 solid horizontal lines show abundance ratios predicted by theoretical
models by WW95 with $Z$ = 0 and $Z$ = 0.01 $Z_\odot$, in
the case of C production by high-mass stars only. The C/O ratio predicted in
the case of C production by both high and intermediate-mass stars is shown
by the horizontal dashed line. The stellar yields for the latter are from
RV81. For N/O (Figure 2b), the solid line shows the mean
\underline{observed} value for high-mass stars, as there is no model for 
primary N production by high-mass stars as yet. The dashed line shows the
expected N/O when primary N is produced by both high and intermediate-mass
stars. The stellar yields for the latter are from RV81. The 
dotted lines in Figures 2a and 2b show the solar ratios. }

\figcaption[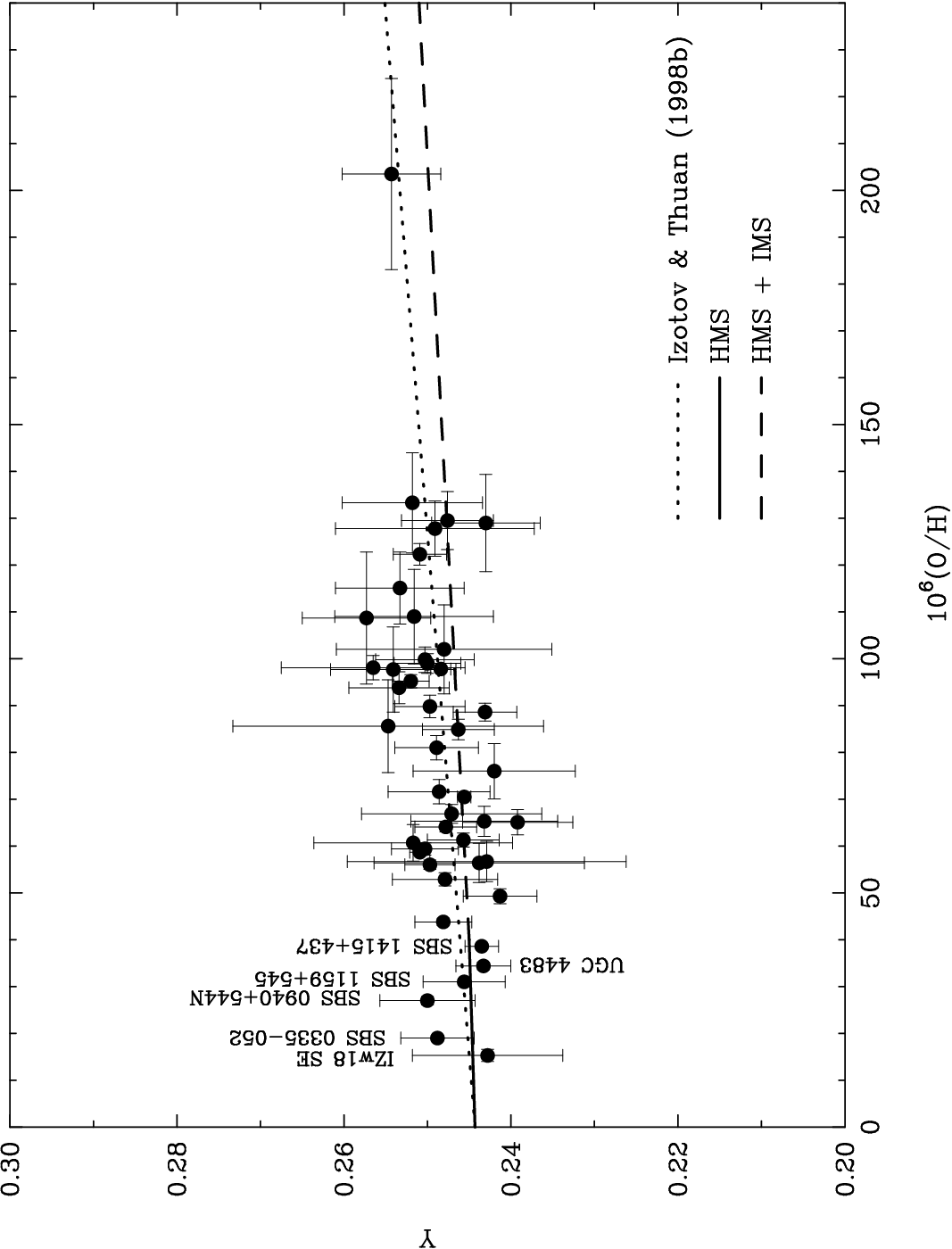]{ The helium mass fraction $Y$ versus oxygen abundance for 
BCGs in the sample of Izotov \& Thuan (1998b). The solid line represents 
Equation 7 with $Y_p$ = 0.244 (Izotov \& Thuan 1998b) and d$Y$/d(O/H) predicted
by WW95 models in the case where He is made only by high-mass stars.
The dashed line represents Equation 7 with the expected d$Y$/d(O/H) when 
He is produced by both high and intermediate-mass stars. The stellar yields
for the latter are from RV81. The dotted line is the best
fit linear regression line for the whole sample obtained
by Izotov \& Thuan (1998b). }

\figcaption[fig.4.eps] {The helium mass fraction $Y$ versus 
the carbon abundance C/H 
for 4 BCGs from the Izotov \& Thuan (1998b) sample. The theoretical dependences
are shown by the solid line for helium and carbon enrichment by high-mass
stars (HMS) only and by the dashed line when He and C are produced by both high
and intermediate-mass stars (HMS + IMS). The theoretical yields for
HMS are IMF-averaged yields for massive star models with 
$Z$ = 10$^{-4}$ $Z_\odot$ (WW95). The theoretical
yields for intermediate-mass stars are IMF-averaged yields from
RV81 for the models with heavy element
mass fraction $Z$ = 0.004 and mixing scale parameter $\alpha$ = 1.5.
The Salpeter slope --2.35 and lower and higher mass limits
of 1 $M_\odot$ and 100 $M_\odot$ respectively have been adopted for the IMF.}

\figcaption[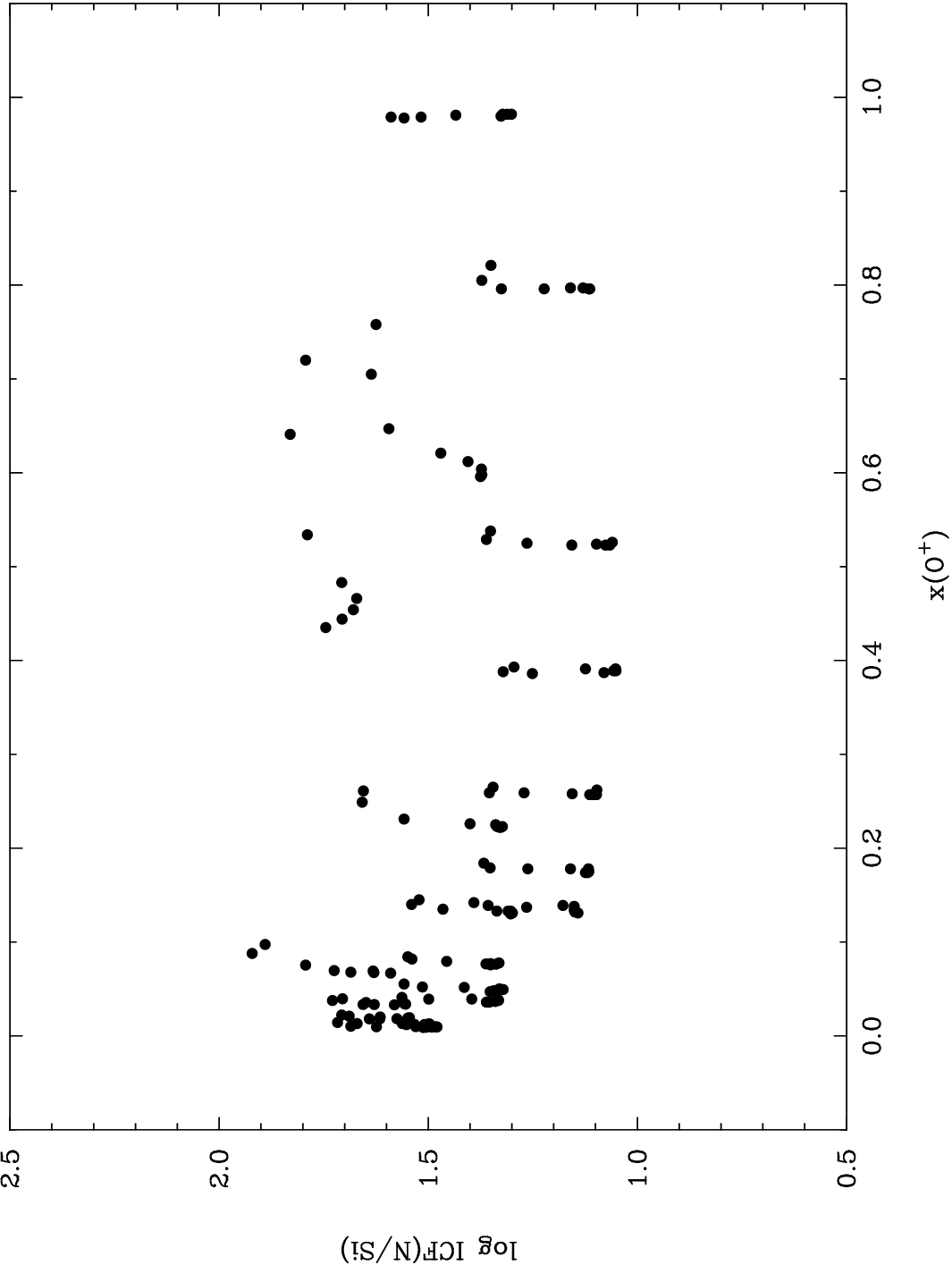] {The correction factor ICF(N/Si) as defined in Equation 9
versus the ion fraction $x$(O$^+$) = O$^+$/O 
as derived from photoionized H II region models by Stasi\'nska (1990). }

\clearpage

%
%
\begin{deluxetable}{lcccccclc}
\footnotesize
\tablenum{1}
\tablecolumns{9}
\tablewidth{0pt}
\tablecaption{Heavy Element to Oxygen Abundance Ratios from Optical 
Observations}
\tablehead{
\colhead{Object}& \colhead{12 + log O/H}&\colhead{log N/O}&\colhead{log Ne/O}
&\colhead{log S/O}&\colhead{log Ar/O}&\colhead{[O/Fe]}&Other name&Ref. }
\startdata
0112--011 & 
8.31$\pm$0.04&--1.26$\pm$0.10&--0.79$\pm$0.08&--1.60$\pm$0.06&--2.25$\pm$0.05& 
0.74$\pm$0.18&UM 311& 1 \nl
0218+003  & 7.93$\pm$0.05&--1.08$\pm$0.12&--0.74$\pm$0.10& \nodata       
&--2.07$\pm$0.06& 0.26$\pm$0.17&UM 420& 1 \nl
0248+042  & 
7.83$\pm$0.01&--1.67$\pm$0.03&--0.77$\pm$0.02&--1.50$\pm$0.02&--2.20$\pm$0.02& 
0.17$\pm$0.02&Mrk 600& 1 \nl
0335--052 & 
7.29$\pm$0.01&--1.58$\pm$0.03&--0.80$\pm$0.03&--1.59$\pm$0.04&--2.14$\pm$0.04&--
0.11$\pm$0.06~\,&& 1 \nl
0459--043 & 
8.04$\pm$0.06&--1.05$\pm$0.14&--0.75$\pm$0.12&--1.66$\pm$0.12&--2.20$\pm$0.07& 
0.47$\pm$0.17&Mrk 1089& 1 \nl
0635+756  & 
8.04$\pm$0.04&--1.36$\pm$0.10&--0.84$\pm$0.08&--1.48$\pm$0.08&--2.19$\pm$0.05& 
0.14$\pm$0.07&Mrk 5& 1 \nl
0723+692A & 
7.85$\pm$0.01&--1.52$\pm$0.01&--0.78$\pm$0.01&--1.48$\pm$0.01&--2.22$\pm$0.01& 
0.49$\pm$0.02&Mrk 71& 2 \nl
0723+692B & 
7.81$\pm$0.02&--1.57$\pm$0.04&--0.77$\pm$0.04&--1.48$\pm$0.04&--2.19$\pm$0.02& 
\nodata      &Mrk 71& 2 \nl
0741+535  & 
8.01$\pm$0.04&--1.54$\pm$0.10&--0.70$\pm$0.08&--1.58$\pm$0.08&--2.30$\pm$0.05& 
0.47$\pm$0.16&Mrk 1409& 2 \nl
0749+568  & 
7.85$\pm$0.05&--1.44$\pm$0.11&--0.72$\pm$0.09&--1.54$\pm$0.10&--2.22$\pm$0.06& 
\nodata      && 2 \nl
0749+582  & 
8.13$\pm$0.03&--1.36$\pm$0.06&--0.62$\pm$0.05&--1.64$\pm$0.06&--2.36$\pm$0.04& 
0.41$\pm$0.13&& 2 \nl
0832+699  & 
7.54$\pm$0.01&--1.64$\pm$0.03&--0.74$\pm$0.02&--1.57$\pm$0.02&--2.25$\pm$0.02& 
\nodata      &UGC 4483& 3 \nl
0907+527  & 
7.97$\pm$0.03&--1.58$\pm$0.07&--0.64$\pm$0.06&--1.57$\pm$0.08&--2.25$\pm$0.08& 
\nodata      && 2 \nl
0917+527  & 
7.86$\pm$0.02&--1.62$\pm$0.04&--0.62$\pm$0.03&--1.62$\pm$0.04&--2.37$\pm$0.02& 
0.47$\pm$0.10&Mrk 1416& 2 \nl
0926+606  & 
7.91$\pm$0.01&--1.47$\pm$0.03&--0.67$\pm$0.03&--1.53$\pm$0.03&--2.35$\pm$0.05& 
0.41$\pm$0.05&& 2 \nl
0930+554  & 
7.18$\pm$0.01&--1.60$\pm$0.03&--0.78$\pm$0.03&--1.65$\pm$0.05&--2.30$\pm$0.03& 
0.38$\pm$0.12&I Zw 18& 2 \nl
0940+544N & 
7.43$\pm$0.01&--1.61$\pm$0.03&--0.72$\pm$0.03&--1.61$\pm$0.03&--2.22$\pm$0.03& 
0.34$\pm$0.04&& 2 \nl
0943+561A & 7.74$\pm$0.06& \nodata       
&--0.64$\pm$0.12&--1.46$\pm$0.15&--2.08$\pm$0.10& \nodata      && 2 \nl
0946+558  & 
8.00$\pm$0.01&--1.51$\pm$0.02&--0.69$\pm$0.02&--1.49$\pm$0.02&--2.34$\pm$0.02& 
0.57$\pm$0.03&Mrk 22& 3 \nl
0948+532  & 
8.00$\pm$0.01&--1.45$\pm$0.03&--0.69$\pm$0.02&--1.54$\pm$0.02&--2.15$\pm$0.02& 
0.33$\pm$0.02&& 3 \nl
1030+583  & 
7.79$\pm$0.01&--1.60$\pm$0.02&--0.77$\pm$0.02&--1.46$\pm$0.02&--2.22$\pm$0.04& 
0.42$\pm$0.07&Mrk 1434& 2 \nl
1053+064  & 
7.99$\pm$0.01&--1.38$\pm$0.03&--0.67$\pm$0.02&--1.62$\pm$0.04&--2.28$\pm$0.04& 
0.34$\pm$0.07&Mrk 1271& 1 \nl
1054+365  & 
7.97$\pm$0.02&--1.48$\pm$0.03&--0.71$\pm$0.03&--1.44$\pm$0.03&--2.23$\pm$0.04& 
\nodata      &CG 798& 2 \nl
1102+294  & 
7.81$\pm$0.02&--1.49$\pm$0.05&--0.76$\pm$0.04&--1.53$\pm$0.06&--2.17$\pm$0.03& 
0.37$\pm$0.11&Mrk 36& 1 \nl
1102+450  & 8.12$\pm$0.03&--1.39$\pm$0.08&--0.61$\pm$0.07& \nodata       
&--2.41$\pm$0.04& 0.50$\pm$0.12&Mrk 162& 1 \nl
1116+583B & 
7.68$\pm$0.05&--1.45$\pm$0.11&--0.78$\pm$0.09&--1.46$\pm$0.14&--2.06$\pm$0.07& 
\nodata      && 2 \nl
1124+792  & 
7.69$\pm$0.01&--1.53$\pm$0.03&--0.81$\pm$0.03&--1.58$\pm$0.03&--2.10$\pm$0.02& 
\nodata      &VII Zw 403& 2 \nl
1128+573  & 7.75$\pm$0.03&--1.51$\pm$0.07&--0.77$\pm$0.06& \nodata       
&--2.31$\pm$0.05& \nodata      && 2 \nl
1135+581  & 
7.98$\pm$0.01&--1.45$\pm$0.01&--0.64$\pm$0.01&--1.60$\pm$0.01&--2.24$\pm$0.01& 
0.44$\pm$0.02&Mrk 1450& 3 \nl
1139+006  & 
7.99$\pm$0.04&--1.01$\pm$0.10&--0.76$\pm$0.08&--1.67$\pm$0.08&--2.30$\pm$0.05& 
0.59$\pm$0.17&UM 448& 1 \nl
1147+153  & 
8.11$\pm$0.02&--1.45$\pm$0.05&--0.74$\pm$0.04&--1.51$\pm$0.04&--2.17$\pm$0.02& 
0.50$\pm$0.12&Mrk 750& 1 \nl
1148--020 & 
7.78$\pm$0.03&--1.50$\pm$0.06&--0.86$\pm$0.05&--1.51$\pm$0.08&--2.27$\pm$0.08& 
\nodata      &UM 461& 1 \nl
1150--021 & 
7.95$\pm$0.01&--1.51$\pm$0.03&--0.76$\pm$0.02&--1.57$\pm$0.03&--2.27$\pm$0.03& 
0.54$\pm$0.07&UM 462& 1 \nl
1152+579  & 
7.81$\pm$0.01&--1.45$\pm$0.02&--0.71$\pm$0.01&--1.57$\pm$0.01&--2.27$\pm$0.01& 
0.28$\pm$0.02&Mrk 193& 3 \nl
1159+545  & 
7.49$\pm$0.01&--1.58$\pm$0.04&--0.73$\pm$0.04&--1.51$\pm$0.04&--2.13$\pm$0.05& 
0.14$\pm$0.05&& 1 \nl
1205+557  & 
7.75$\pm$0.03&--1.50$\pm$0.08&--0.68$\pm$0.06&--1.61$\pm$0.08&--2.35$\pm$0.05& 
\nodata      && 2 \nl
1211+540  & 
7.64$\pm$0.01&--1.59$\pm$0.02&--0.75$\pm$0.02&--1.48$\pm$0.02&--2.39$\pm$0.02& 
0.47$\pm$0.02&& 3 \nl
1222+614  & 
7.95$\pm$0.01&--1.61$\pm$0.02&--0.71$\pm$0.02&--1.56$\pm$0.02&--2.32$\pm$0.03& 
0.41$\pm$0.06&& 2 \nl
1223+487  & 
7.77$\pm$0.01&--1.49$\pm$0.01&--0.75$\pm$0.01&--1.47$\pm$0.01&--2.22$\pm$0.01& 
0.49$\pm$0.03&Mrk 209& 2 \nl
1249+493  & 
7.72$\pm$0.01&--1.59$\pm$0.03&--0.68$\pm$0.02&--1.66$\pm$0.03&--2.40$\pm$0.02& 
\nodata      && 3 \nl
1256+351  & 
7.99$\pm$0.01&--1.52$\pm$0.01&--0.74$\pm$0.01&--1.54$\pm$0.01&--2.33$\pm$0.01& 
0.50$\pm$0.02&Mrk 59& 2 \nl
1319+579A & 
8.11$\pm$0.03&--1.40$\pm$0.08&--0.64$\pm$0.07&--1.60$\pm$0.06&--2.41$\pm$0.04& 
0.49$\pm$0.12&& 2 \nl
1319+579B & 
8.09$\pm$0.01&--1.47$\pm$0.02&--0.76$\pm$0.02&--1.58$\pm$0.02&--2.40$\pm$0.03& 
0.45$\pm$0.04&& 2 \nl
1331+493N & 
7.77$\pm$0.01&--1.55$\pm$0.02&--0.71$\pm$0.02&--1.53$\pm$0.02&--2.46$\pm$0.03& 
0.34$\pm$0.03&& 3 \nl
1331+493S & 7.87$\pm$0.03&--1.50$\pm$0.09&--0.66$\pm$0.07& \nodata       & 
\nodata       & \nodata      && 4 \nl
1358+576  & 
7.88$\pm$0.01&--1.31$\pm$0.03&--0.70$\pm$0.03&--1.61$\pm$0.04&--2.35$\pm$0.06& 
0.31$\pm$0.05&Mrk 1486& 2 \nl
1415+437  & 
7.59$\pm$0.01&--1.58$\pm$0.01&--0.73$\pm$0.01&--1.59$\pm$0.01&--2.29$\pm$0.01& 
0.41$\pm$0.02&CG 389& 1 \nl
1420+544  & 
7.75$\pm$0.01&--1.56$\pm$0.02&--0.75$\pm$0.01&--1.53$\pm$0.01&--2.18$\pm$0.01& 
\nodata      &CG 413& 1 \nl
1437+370  & 
7.93$\pm$0.01&--1.50$\pm$0.03&--0.69$\pm$0.02&--1.52$\pm$0.02&--2.18$\pm$0.02& 
0.42$\pm$0.02&Mrk 475& 3 \nl
1441+294  & 7.99$\pm$0.06&--1.41$\pm$0.13&--0.78$\pm$0.11& \nodata       
&--2.30$\pm$0.14& \nodata      &CG 1258& 2 \nl
1533+574A & 
7.88$\pm$0.03&--1.43$\pm$0.08&--0.68$\pm$0.07&--1.55$\pm$0.07&--2.20$\pm$0.04& 
0.34$\pm$0.12&& 2 \nl
1533+574B & 
8.11$\pm$0.02&--1.54$\pm$0.05&--0.71$\pm$0.04&--1.62$\pm$0.04&--2.29$\pm$0.06& 
0.40$\pm$0.06&& 2 \nl
1535+554  & 
8.06$\pm$0.04&--1.50$\pm$0.10&--0.69$\pm$0.08&--1.53$\pm$0.09&--2.30$\pm$0.06& 
\nodata      &Mrk 487& 2 \nl
2329+286  & 
8.06$\pm$0.03&--1.39$\pm$0.07&--0.71$\pm$0.06&--1.65$\pm$0.08&--2.33$\pm$0.04& 
0.34$\pm$0.10&Mrk 930& 1 \nl
\enddata
\tablerefs{1 - Izotov \& Thuan 1998b; 2 - ITL97; 3 - ITL94; 4 - TIL95.}
\end{deluxetable}

\clearpage

%
%

\begin{deluxetable}{lcccccccc}
\small
\tablenum{2}
\tablecolumns{9}
\tablewidth{0pt}
\tablecaption{Emission Line Intensities from the MMT spectra}
\tablehead{
\colhead{}& \multicolumn{2}{c}{I Zw 18 NW}&\colhead{}&\multicolumn{2}{c}{I Zw 18 
SE}&\colhead{}&\multicolumn{2}{c}{SBS 0335--052} \nl 
\cline{2-3} \cline{5-6} \cline{8-9} \nl
\colhead{}&\colhead{$F$($\lambda$)/$F$(H$\beta$)}
&\colhead{$I$($\lambda$)/$I$(H$\beta$)}&\colhead{}
&\colhead{$F$($\lambda$)/$F$(H$\beta$)}
&\colhead{$I$($\lambda$)/$I$(H$\beta$)}&\colhead{}
&\colhead{$F$($\lambda$)/$F$(H$\beta$)}
&\colhead{$I$($\lambda$)/$I$(H$\beta$)} }
\startdata
 1883 + 1892\ Si III]&\nodata        & \nodata        &&\nodata        &\nodata  
      &&0.121$\pm$0.010&0.198$\pm$0.020 \nl
 1907\ C III]        &\nodata        & \nodata        &&\nodata        
&\nodata        &&0.300$\pm$0.015&0.490$\pm$0.025 \nl
 3727\ [O II]        &0.153$\pm$0.007& 
0.158$\pm$0.008&&0.497$\pm$0.008&0.489$\pm$0.008&&0.171$\pm$0.001
&0.183$\pm$0.002 \nl
 3835\ H9            &\nodata        & \nodata        
&&0.041$\pm$0.004&0.087$\pm$0.011&&0.043$\pm$0.001&0.082$\pm$0.002 \nl
 3868\ [Ne III]      &0.138$\pm$0.007& 
0.140$\pm$0.008&&0.152$\pm$0.005&0.149$\pm$0.005&&0.179$\pm$0.001
&0.189$\pm$0.002 \nl
 3889\ He I + H8     &0.083$\pm$0.007& 
0.208$\pm$0.023&&0.158$\pm$0.005&0.200$\pm$0.007&&0.122$\pm$0.001
&0.163$\pm$0.002 \nl
 3968\ [Ne III] + H7 &0.094$\pm$0.007& 
0.215$\pm$0.021&&0.172$\pm$0.005&0.213$\pm$0.007&&0.175$\pm$0.001
&0.217$\pm$0.002 \nl
 4101\ H$\delta$     &0.177$\pm$0.007& 
0.284$\pm$0.015&&0.231$\pm$0.005&0.267$\pm$0.007&&0.203$\pm$0.001
&0.242$\pm$0.002 \nl
 4340\ H$\gamma$     &0.383$\pm$0.008& 
0.467$\pm$0.012&&0.450$\pm$0.007&0.474$\pm$0.008&&0.418$\pm$0.002
&0.453$\pm$0.002 \nl
 4363\ [O III]       &0.078$\pm$0.007& 
0.076$\pm$0.008&&0.059$\pm$0.004&0.058$\pm$0.004&&0.096$\pm$0.001
&0.097$\pm$0.001 \nl
 4471\ He I          &0.012$\pm$0.005& 
0.011$\pm$0.006&&0.034$\pm$0.004&0.033$\pm$0.004&&0.031$\pm$0.001& 0.033$\pm$0.001 \nl
 4686\ He II         &0.038$\pm$0.006& 0.036$\pm$0.006&&\nodata        
&\nodata        &&0.020$\pm$0.001& 0.020$\pm$0.001 \nl
 4861\ H$\beta$      &1.000$\pm$0.012& 
1.000$\pm$0.014&&1.000$\pm$0.011&1.000$\pm$0.012&&1.000$\pm$0.004
&1.000$\pm$0.004 \nl
 4959\ [O III]       &0.729$\pm$0.010& 
0.675$\pm$0.010&&0.620$\pm$0.008&0.603$\pm$0.008&&1.056$\pm$0.004
&1.026$\pm$0.004 \nl
 5007\ [O III]       &2.173$\pm$0.022& 
2.003$\pm$0.022&&1.850$\pm$0.018&1.799$\pm$0.018&&3.163$\pm$0.009
&3.061$\pm$0.009 \nl
 5876\ He I          &0.063$\pm$0.004& 
0.055$\pm$0.004&&0.095$\pm$0.003&0.092$\pm$0.003&&0.115$\pm$0.001
&0.105$\pm$0.001 \nl
 6563\ H$\alpha$     &3.235$\pm$0.030& 
2.727$\pm$0.030&&2.832$\pm$0.026&2.738$\pm$0.028&&3.091$\pm$0.009
&2.747$\pm$0.009 \nl
 6678\ He I          &0.027$\pm$0.004& 
0.023$\pm$0.003&&0.030$\pm$0.002&0.029$\pm$0.002&&0.032$\pm$0.001
&0.028$\pm$0.001 \nl
 6717\ [S II]        &0.021$\pm$0.004& 
0.017$\pm$0.003&&0.043$\pm$0.003&0.042$\pm$0.003&&0.023$\pm$0.001
&0.020$\pm$0.001 \nl
 6731\ [S II]        &0.017$\pm$0.004& 
0.014$\pm$0.003&&0.029$\pm$0.002&0.028$\pm$0.002&&0.022$\pm$0.001
&0.019$\pm$0.001 \nl \nl
 $C$(H$\beta$) dex    &\multicolumn {2}{c}{0.145}&\colhead{}&\multicolumn 
{2}{c}{0.015}&\colhead{}&\multicolumn {2}{c}{0.130} \nl
 $F$(H$\beta$)\tablenotemark{a} &\multicolumn {2}{c}{ 
0.29}&\colhead{}&\multicolumn{2}{c}{0.35}&\colhead{}&\multicolumn{2}{c}{1.91} 
\nl
 $EW$(H$\beta$)\ \AA &\multicolumn {2}{c}{34}&\colhead{}&\multicolumn 
{2}{c}{144}&\colhead{}&\multicolumn {2}{c}{237} \nl
 $EW$(abs)\ \AA      &\multicolumn {2}{c}{2.5}&\colhead{}&\multicolumn{2}{c}{ 
3.9}&\colhead{}&\multicolumn{2}{c}{ 5.4} \nl
\enddata
\tablenotetext{a}{in units of 10$^{-14}$ ergs\ s$^{-1}$cm$^{-2}$.}
\end{deluxetable}

\clearpage

\begin{deluxetable}{lccc}
\tablenum{3}
\tablecolumns{4}
\tablewidth{400pt}
\tablecaption{Electron Temperatures and Oxygen Abundance Derived from the MMT 
Spectra}
\tablehead{
\colhead{Property}&\colhead{I Zw 18 NW}
&\colhead{I Zw 18 SE}&\colhead{SBS 0335--052} }
\startdata
$T_e$(O III)(K)                     &21,500$\pm$1,400      &19,500$\pm$800      
&19,300$\pm$100      \nl
$T_e$(O II)(K)                      &16,100$\pm$~~900      &15,500$\pm$600      
&15,500$\pm$100      \nl
$T_e$(S III)(K)                     &19,500$\pm$1,100      &17,900$\pm$600      
&17,700$\pm$100      \nl
O$^+$/H$^+$($\times$10$^5$)         &0.11~~$\pm$~0.02      &0.38~~$\pm$~0.04    
&0.15~~$\pm$~0.01      \nl
O$^{++}$/H$^+$($\times$10$^5$)      &1.00~~$\pm$~0.14      &1.09~~$\pm$~0.10    
&1.90~~$\pm$~0.01      \nl
O$^{3+}$/H$^+$($\times$10$^5$)      &0.05~~$\pm$~0.01      &\nodata    
&0.04~~$\pm$~0.01      \nl
O/H($\times$10$^5$)                 &1.17~~$\pm$~0.15      &1.48~~$\pm$~0.10    
&2.09~~$\pm$~0.01      \nl
12 + log(O/H)                       &7.07~~$\pm$~0.05      &7.17~~$\pm$~0.06    
&7.32~~$\pm$~0.01      \nl
\enddata
\end{deluxetable}

\clearpage

%
%

\begin{figure}
\epsscale{1.6}
\plotfiddle{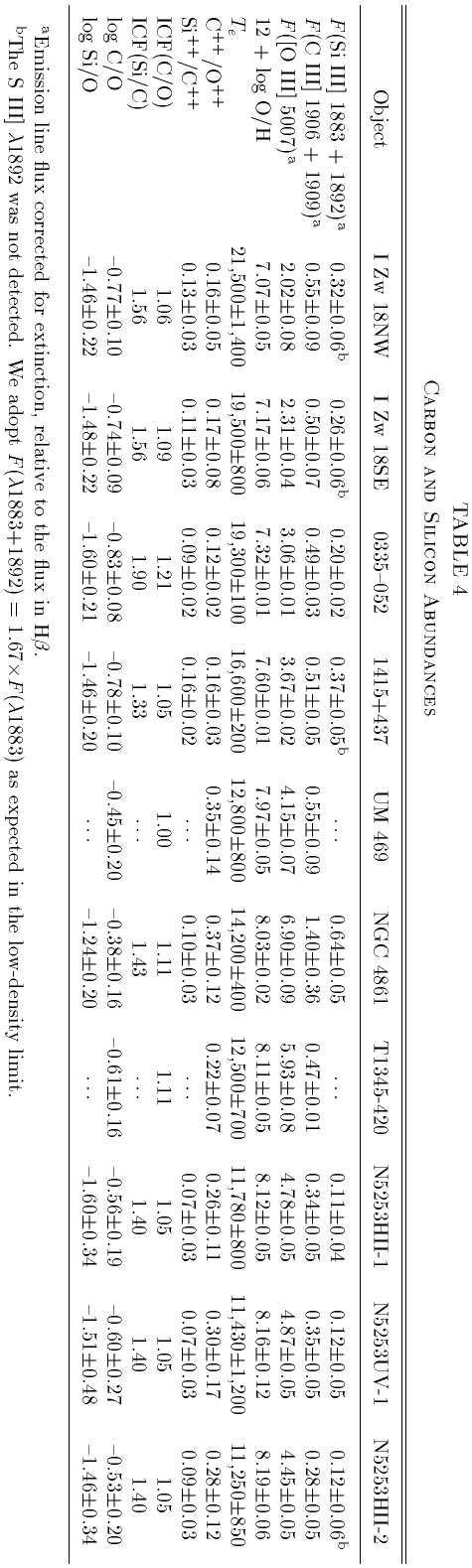}{0.cm}{180.}{90.}{90.}{330.}{390.}
\end{figure}

\clearpage

\begin{deluxetable}{lrrrr}
\tablenum{5}
\tablecolumns{5}
\tablewidth{0pt}
\tablecaption{Mean Heavy Element Abundance Ratios}
\tablehead{
\colhead{Quantity}&\colhead{Total sample}
&\colhead{Low-metallicity}&\colhead{High-metallicity}
&\colhead{Sun\tablenotemark{c}} \nl
\colhead{}&\colhead{}
&\colhead{subsample\tablenotemark{a}}&\colhead{subsample\tablenotemark{b}}
&\colhead{} }
\startdata
log C/O    &--0.63$\pm$0.14 (10)\tablenotemark{d}  &--0.78$\pm$0.03 
(~4)&--0.52$\pm$0.08 (\,~6)&--0.38 \nl
log N/O    &--1.47$\pm$0.14 (53)  &--1.60$\pm$0.02 (~6)&--1.46$\pm$0.14 (47)  
&--0.88 \nl
log Ne/O   &--0.72$\pm$0.06 (54)  &--0.75$\pm$0.03 (~6)&--0.72$\pm$0.06 (48)  
&--0.84 \nl
log Si/O   &--1.48$\pm$0.11 (\,~8)&--1.50$\pm$0.06 (~4)&--1.45$\pm$0.13 
(\,~4)&--1.37 \nl
log S/O    &--1.56$\pm$0.06 (49)  &--1.59$\pm$0.04 (~6)&--1.55$\pm$0.06 (43)  
&--1.66 \nl
log Ar/O   &--2.26$\pm$0.09 (53)  &--2.22$\pm$0.07 (~6)&--2.27$\pm$0.10 (47)  
&--2.37 \nl
[O/Fe]     &  0.40$\pm$0.14 (38)  &  0.32$\pm$0.11 (~4)\tablenotemark{e}&  
0.42$\pm$0.12 (33)  &  0.00 \nl
\enddata
\tablenotetext{a}{Galaxies with 12 + log O/H $\leq$ 7.6.}
\tablenotetext{b}{Galaxies with 12 + log O/H $>$ 7.6.}
\tablenotetext{c}{Solar abundance ratios are from Anders \& Grevesse
(1989).}
\tablenotetext{d}{The number of H II regions entering in the mean 
is shown in parentheses.}
\tablenotetext{e}{The SBS 0335--052 deviant point is excluded. Its 
inclusion would have given $<$[O/Fe]$>$ = 0.23$\pm$0.20.}
\end{deluxetable}

\clearpage

\begin{deluxetable}{lcccccc}
\tablenum{6}
\tablecolumns{7}
\tablewidth{0pt}
\tablecaption{Heavy Element Yields from Massive Stars}
\tablehead{
\colhead{}&\colhead{}&\multicolumn{5}{c}{Models\tablenotemark{a}} \nl 
\cline{3-7}
\colhead{Quantity}&\colhead{Observations}
&\colhead{$Z$=0}&\colhead{$Z$=10$^{-4}$$Z_\odot$} 
&\colhead{$Z$=10$^{-2}$$Z_\odot$}&\colhead{$Z$=10$^{-1}$$Z_\odot$}&\colhead{$Z$=
$Z_\odot$} }
\startdata
$M$(O), $M_\odot$ &\nodata                   & 3.59E--2& 3.59E--2& 3.78E--2& 
3.96E--2& 4.48E--2 \nl
$M$(C)/$M$(O)     & 1.24E--1\tablenotemark{b}& 1.39E--1& 1.34E--1& 1.22E--1& 
1.16E--1& 1.17E--1 \nl
$M$(N)/$M$(O)     & 2.19E--2\tablenotemark{b}& 2.64E--4& 4.67E--5& 4.77E--4& 
3.77E--3& 3.82E--2 \nl
$M$(Ne)/$M$(O)    & 2.38E--1\tablenotemark{c}& 2.48E--1& 1.08E--1& 9.73E--2& 
1.11E--1& 1.33E--1 \nl
$M$(Si)/$M$(O)    & 5.79E--2\tablenotemark{c}& 7.19E--2& 9.98E--2& 1.02E--1& 
1.01E--1& 1.15E--1 \nl
$M$(S)/$M$(O)     & 5.51E--2\tablenotemark{c}& 3.56E--2& 4.69E--2& 4.89E--2& 
4.83E--2& 5.41E--2 \nl
$M$(Ar)/$M$(O)    & 1.23E--2\tablenotemark{c}& 7.32E--3& 9.54E--3& 1.05E--2& 
1.01E--2& 1.12E--2 \nl
$M$(Fe)/$M$(O)    & 5.55E--2\tablenotemark{c}& 1.31E--1& 9.99E--2& 1.17E--1& 
1.18E--1& 9.55E--2 \nl
\enddata
\tablenotetext{a}{IMF-averaged yield ratios. The Salpeter IMF with slope
$\alpha$=--2.35 is adopted. The lower and upper mass limits are 1 $M_\odot$ and
100 $M_\odot$ respectively. The stellar yields are from WW95.}
\tablenotetext{b}{Mean value for the BCGs with 12 + log O/H $\leq$ 7.6.}
\tablenotetext{c}{Mean value for the total sample.}
\end{deluxetable}

\clearpage

%
%
\begin{deluxetable}{lcccccccccc}
\small
\tablenum{7}
\tablecolumns{11}
\tablewidth{0pt}
\tablecaption{Comparison of Observed and Theoretical Abundance Ratios}
\tablehead{
\colhead{}&\multicolumn{2}{c}{}&&\multicolumn{7}{c}{Models\tablenotemark{a}} \nl 
\cline{5-11}
\colhead{}&\multicolumn{2}{c}{Observations}&&\multicolumn{3}{c}{RV81}&
&\multicolumn{3}{c}{HG97} \nl 
\cline{2-3} \cline{5-7} \cline{9-11}
\colhead{Quantity}&\colhead{12 +logO/H$\leq$7.6}&\colhead{12 +logO/H$>$7.6}
&&\colhead{IMS}&\colhead{HMS}&\colhead{IMS + 
HMS}&&\colhead{IMS}&\colhead{HMS}&\colhead{IMS + HMS} }
\startdata
log(C/O)     & --0.78$\pm$0.03& --0.52$\pm$0.08&&\nodata& --0.75& --0.31&&\nodata& --0.75& --0.55 \nl
log(N/O)     & --1.60$\pm$0.02& --1.46$\pm$0.14&& --1.27& --1.60\tablenotemark{b}& --1.11&& --0.84& --1.60\tablenotemark{b}& --0.77 \nl
d$Y$/d(O/H)  & \multicolumn{2}{c}{45$\pm$19} &&\nodata&17&30&&\nodata&17&28 \nl
d$Y$/d(C/H)  & \multicolumn{2}{c}{\nodata} &&\nodata&129&83&&\nodata&129&133 \nl
d$Y$/d$Z$    & \multicolumn{2}{c}{2.4$\pm$1.0} &&\nodata&~\,0.94&~\,1.66&&\nodata&~\,0.94&~\,1.54 \nl
\enddata
\tablenotetext{a}{IMF-averaged abundance ratios. The Salpeter IMF with slope
$\alpha$=--2.35 is adopted. The lower and upper mass limits are 1 $M_\odot$ and
100 $M_\odot$ respectively. The yields for the high-mass stars (HMS) 
with $Z$=0.0002 are from WW95.
The yields for the intermediate-mass stars (IMS)  
are from RV81 for $Z$=0.004 and a mixing length 
parameter $\alpha$=1.5 and from HG97 for $Z$=0.001 and a mass loss 
efficiency parameter $\eta$=1. In the latter case, stellar masses range from 
2 to 7 $M_\odot$. }
\tablenotetext{b}{Value adopted from the observations.}
\end{deluxetable}

\clearpage
%
%
\begin{figure}
\figurenum{1}
\epsscale{1.6}
\plotfiddle{fig1.eps}{0.cm}{0.}{90.}{90.}{-130.}{-300.}
\end{figure}

\clearpage
%
%
\begin{figure*}
\figurenum{2}
\epsscale{2.0}
\plotfiddle{fig2.eps}{0.cm}{0.}{90.}{90.}{-260.}{-330.}
\end{figure*}

\clearpage
%
%
\begin{figure*}
\figurenum{3}
\epsscale{2.0}
\plotfiddle{fig3.eps}{0.cm}{0.}{88.}{88.}{-270.}{-270.}
\end{figure*}

\clearpage
%
%
\begin{figure*}
\figurenum{4}
\epsscale{2.0}
\plotfiddle{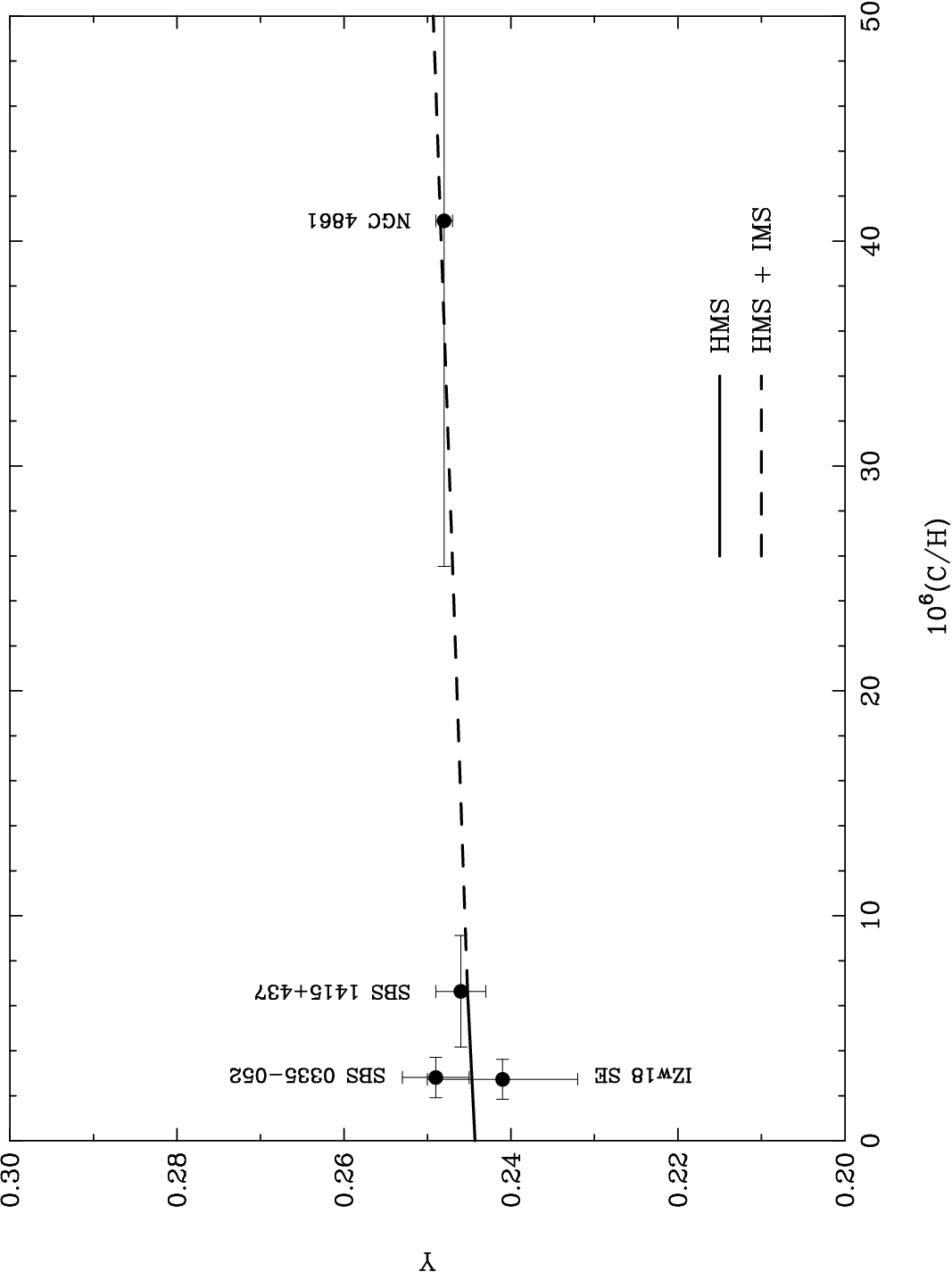}{0.cm}{0.}{88.}{88.}{-270.}{-270.}
\end{figure*}

\clearpage
%
%
\begin{figure*}
\figurenum{5}
\epsscale{2.0}
\plotfiddle{fig5.eps}{0.cm}{0.}{88.}{88.}{-270.}{-270.}
\end{figure*}

\end{document}